\RequirePackage{fix-cm}
\documentclass[twocolumn]{svjour3}          % twocolumn
\smartqed  % flush right qed marks, e.g. at end of proof

\usepackage{graphicx}      % include this line if your document contains figures
\usepackage{cite}        % required for bibliography
\usepackage{siunitx}
\usepackage{amsmath,cancel}
\usepackage{array}
\usepackage{amsmath}
\usepackage{tikz}
\usepackage{mathdots}
\usepackage{yhmath}
\usepackage{cancel}
\usepackage{color}
\usepackage{siunitx}
\usepackage{array}
\usepackage{multirow}
\usepackage{amssymb}
\usepackage{gensymb}
\usepackage{tabularx}
\usepackage{booktabs}
\usepackage{pgfplots}
\usepackage{subcaption}
\usepackage{color, soul}
\sethlcolor{yellow}
\pgfplotsset{compat=newest}
\usetikzlibrary{plotmarks}
\usetikzlibrary{arrows.meta}
\usepgfplotslibrary{patchplots}
\usepackage{grffile}
\usepackage{amsmath}
\usetikzlibrary{fadings}
\usetikzlibrary{patterns}
\renewcommand{\hl}{}
\begin{document}
	
	\title{Fractional-Order Single State Reset Element\thanks{This work was supported by NWO, through OTP TTW project \#16335}
	}
	
	
	\author{Nima Karbasizadeh         \and
		Niranjan Saikumar \and
		S. Hassan HosseinNia %etc.
	}
	
	
	\institute{Nima Karbasizadeh    \&
		Niranjan Saikumar \&
		S. Hassan HosseinNia  \at
		Department of Precision and Microsystem Engineering, Delft University of Technology, Delft, The Netherlands \\\\
		N. Karbasizadeh,
		\email{n.karbasizadehesfahani@tudelft.nl}\\
		N. Saikumar,
		\email{n.saikumar@tudelft.nl}\\
		S. H. HosseinNia,
		\email{s.h.hosseinnia@tudelft.nl}           %  \\
	}
	
	\date{Received: date / Accepted: date}

	\maketitle
	
	\begin{abstract}
		This paper proposes a fractional-order reset element whose architecture allows for the suppression of nonlinear effects for a range of frequencies. Suppressing the nonlinear effects of a reset element for the desired frequency range while maintaining it for the rest is beneficial, especially when it is used in the framework of a ``Constant in gain, Lead in phase'' (CgLp) filter. CgLp is a newly introduced nonlinear filter, bound to circumvent the well-known linear control limitation -- the waterbed effect. The ideal behaviour of such a filter in the frequency domain is unity gain while providing a phase lead for a broad range of frequencies. However, CgLp's ideal behaviour is based on the describing function, which is a first-order approximation that neglects the effects of the higher-order harmonics in the output of the filter. Although CgLp is fundamentally a nonlinear filter, its nonlinearity is not required for all frequencies. Thus, it is shown in this paper that using the proposed reset element architecture, CgLp gets closer to its ideal behaviour for a range of frequencies, and its performance will be improved accordingly.
	\end{abstract}
	
	\section{Introduction}
	\label{intro}
	PID is still the workhorse of the industry when it comes to the term ``control''. However, in some fields, particularly, precision motion control, there is an increasingly high demand for more precise, faster and more robust controllers. The ``waterbed effect'' is the fundamental and well-known limitation of linear controllers which is preventing them from meeting these demands simultaneously~\cite{maclejowski1989multivariate}. In terms of steady-state precision, increasing the system gain at lower frequencies while decreasing it at higher frequencies will improve the performance of the system. This is according to a well-known frequency-based design method of controllers, known as loop-shaping~\cite{schmidt2014design}. However, according to Bode's gain-phase relationship for linear systems and frequency response of the differentiator part of PID, this desire is in contradiction with the stability of the systems. In other words, improving the performance of the system in terms of precision, speed or stability will be at the cost of deteriorating at least one of the other two.\\
	Nonlinear controllers can be used to circumvent this limitation. Among all of the different types of nonlinearities used by researchers to this end, a relatively simple one was first used by Clegg~\cite{clegg1958nonlinear} in his special integrator. The idea was to reset the value of the integrator whenever its input crosses zero. Clegg showed that his integrator benefits from a phase lead with respect to its linear counterpart. This category of controllers, thereafter called reset controllers, have been further developed and more sophisticated elements such as First Order Reset Element (FORE) in~\cite{horowitz1975non, zaccarian2005first}, Generalised FORE (GFORE) in~\cite{guo2009frequency} and Second-Order Reset Element (SORE) in~\cite{hazeleger2016second}. These reset elements were used in different capacities such as phase lag reduction, decreasing sensitivity peak, narrowband and broadband phase compensation, control of positioning systems with friction see~\cite{wu2006reset, li2005nonlinear,li2010reset,li2011optimal,palanikumar2018no,saikumar2019resetting, BisBee_ARC20a, BeeBis_AUT19a}. Reset elements stability was also investigated in the literature~\cite{LooGru_AUT17a,beker1999stability}.  \\ 
	In almost all of the reset elements studied in the literature, all of the coupled states of the reset element reset~\cite{saikumar2019constant,valerio2019reset,hazeleger2016second}. In other words, there is no coupling between a reset and a non-reset state in the architecture of a reset element. However in~\cite{karbasizadeh2020benefiting}, it was shown that coupling a reset state, and a linear one in an architecture of a SORE can cause the element, called ``Second-Order Single State Reset Element''(SOSRE), to exhibit a linear behaviour in terms of steady-state sinusoidal response at a certain frequency. It was also shown that this phenomenon could be used to the benefit of improving steady-state precision of the overall controller at that certain frequency. However, this architecture cannot be used to suppress higher-order harmonics in a broad range of frequencies.\\
	A new reset-based architecture was recently proposed by~\cite{saikumar2019constant}, which has a constant gain while providing phase lead in a broad range of frequencies. This architecture, named ``Constant in gain, Lead in phase'' (CgLp), can completely replace or take up a significant portion of derivative duties in the framework of PID. In its ideal behaviour, this element robustly stabilises the system by providing the phase lead required in the bandwidth region; however, unlike the derivative part of PID, it does not violate the loop-shaping requirements. Nonetheless, this ideal behaviour is based on the assumptions of the describing function (DF) method, which is a first-order approximation neglecting higher-order harmonics in the output of a nonlinear element. As it will be shown in this paper, DF approximation can be totally unreliable in the cases where the magnitude of higher-order harmonics are relatively large. In some cases, the magnitude of higher-order harmonics can be even larger than first-order one.\\
	Fractional order derivatives and integrals have been used for control designs like Fractional PID~\cite{podlubny1999fractional, dastjerdi2019linear, silva2004fractional, ladaci2006fractional} or CRONE control~\cite{oustaloup1993first,oustaloup1993second,melchior2004motion}. Recently fractional-order elements have also been used within reset elements~\cite{valerio2019reset,hosseinnia2013fractional,valerio2012fractional,saikumar2019complex} in order to approximate complex-order behaviour.\\
	The main contribution of this paper is to use concept of fractional order calculus within reset elements to suppress the nonlinear effects, i.e., higher-order harmonics. As an extension to SOSRE, this paper couples a fractional order integrator with a reset one which creates the ability to selectively  higher-order harmonics, in a range of frequencies where nonlinearity does not have a clear benefit. In the framework of CgLp, nonlinearity is mainly used to create phase lead in the crossover frequency region; however, it is shown that it can have ill effects for other regions, especially, lower frequencies which are important for tracking performance. Moreover, the nonlinear architecture proposed in this paper can be tuned to behave completely linear at a particular frequency which means higher-order harmonics will be zero. Suppressing higher-order harmonics at lower frequencies and eliminating them at a particular frequency will improve the performance of the system in terms of steady-state tracking precision.  \\
	The remainder of this paper is organised as follows. The second section presents the preliminaries. The following one introduces and discusses the architecture of the introduced element. The fourth section will investigate the benefits of the architecture in suppressing the higher-order harmonics at a wide range of frequencies. The following one will introduce an illustrative example and verify the discussions in simulation. Finally, the paper concludes with some remarks and recommendations about ongoing works.
	
	\section{Preliminaries}
	\label{sec:2}
	In this section, the preliminaries of this study will be discussed.
	
	\subsection{General Reset Controller}
	Following is a general form of a rest controller~\cite{Guo:2015}:
	\begin{align}
	\label{eq:reset}
	{{\sum }_{R}}:=\left\{ \begin{aligned}
	& {{{\dot{x}}}_{r}}(t)={{A}}{{x}_{r}}(t)+{{B}}e(t)&\text{if }e(t)\ne 0,\\ 
	& {{x}_{r}}({{t}^{+}})={{A}_{\rho }}{{x}_{r}}(t)&\text{if }e(t)=0, \\ 
	& u(t)={{C}}{{x}_{r}}(t)+{{D}}e(t) \\ 
	\end{aligned} \right.
	\end{align}
	where $A,B,C,D$ are the state space matrices of the base linear system and $A_\rho=\text{diag}(\gamma_1,...,\gamma_n)$ is called reset matrix. This contains the reset coefficients for each state which are denoted by $\gamma_1,...,\gamma_n$. The controller's input and output are represented by $e(t)$ and $ u(t) $, respectively.
	In the spacial case of $A_\rho=I$, no reset will happen and the result is called ``base linear system''.
	\subsection{$H_\beta$ condition}
	The quadratic stability of the closed loop reset system when the base linear system is stable can be examined by the following condition~\cite{beker2004fundamental,guo2015analysis}.
	\begin{theorem}             
		There exists a constant $\beta \in \Re^{n_r\times 1}$ and positive definite matrix $P_\rho \in \Re^{n_r\times n_r}$, such that the restricted Lyapunov equation
		\begin{eqnarray}
		P > 0,\ &A_{cl}^TP + PA_{cl} &< 0\\
		&B_0^TP &= C_0
		\end{eqnarray}
		has a solution for $P$, where $C_0$ and $B_0$ are defined by
		\begin{align}
		C_0=\left[\begin{array}{ccc}
		\beta C_{p} & 0_{n_r \times n_{nr}} & P_\rho
		\end{array}\right] , & &  B_0=\left[\begin{array}{c}
		0_{n_{p} \times n_{r}}\\
		0_{n_{nr} \times n_{r}}\\
		I_{n_r}
		\end{array}\right].
		\end{align}
		And 
		\begin{equation}
		A_{\rho}^TP_\rho A_{\rho} - P_{\rho} \le 0
		\end{equation}
		$A_{cl}$ is the closed loop A-matrix.
		%     \begin{equation}
		%     A_{cl} =\left[\begin{array}{cc}
		%     A_p & B_p C_r\\
		%     -B_rC_p & A_r
		%     \end{array}\right] 
		%     \end{equation}
		%     in which $(A_r,B_r,C_r,D_r)$ are the state space matrices of the controller defined by Eqn. \ref{eq:reset} with 
		$n_r$ is the number of states being reset and $n_{nr}$ being the number of non-resetting states and  ${n_{p}}$ is the number states for the plant.
		$A_p,B_p,C_p,D_p$ are the state space matrices of the plant.
	\end{theorem} 
	\subsection{Describing Functions}
	Because of its nonlinearity, the steady state response of a reset element to a sinusoidal input is not sinusoidal. Thus, its frequency response should be analysed through approximations like Describing Function (DF) method~\cite{guo2009frequency}. However, the DF method only takes the first harmonic of Fourier series decomposition of the output into account and neglects the effects of the higher order harmonics. As shown in~\cite{karbasizadeh2020benefiting}, this simplification can sometimes be significantly inaccurate. To have more accurate  information about the frequency response of nonlinear systems, a method called ``Higher Order Sinusoidal Input Describing Function'' (HOSIDF) has been introduced in~\cite{nuij2006higher}. This method was developed in \cite{kars2018HOSIDF,dastjerdi2020closed} for reset elements defined by~(\ref{eq:reset}) as follows:
	\begin{align}  \nonumber
	& G_n(\omega)=\left\{ \begin{aligned}
	& C{{(j\omega I-A)}^{-1}}(I+j{{\Theta }_{D}}(\omega ))B+D\quad n=1\\ 
	& C{{(j\omega nI-A)}^{-1}}j{{\Theta }_{D}}(\omega )B\qquad\quad\text{odd }n> 2\\ 
	& 0\quad\qquad\qquad\qquad\qquad\qquad\qquad\text{ even }n\ge 2\\ 
	\end{aligned} \right. \\
	&\begin{aligned}
	& {{\Theta }_{D}}(\omega )=-\frac{2{{\omega }^{2}}}{\pi }\Delta (\omega )[{{\Gamma }_{r}}(\omega )-{{\Lambda }^{-1}}(\omega )] \\  
	& \Lambda (\omega )={{\omega }^{2}}I+{{A}^{2}} \\  
	& \Delta (\omega )=I+{{e}^{\frac{\pi }{\omega }A}} \\  
	& {{\Delta }_{r}}(\omega )=I+{{A}_{\rho}}{{e}^{\frac{\pi }{\omega }A}} \\  
	& {{\Gamma }_{r}}(\omega )={{\Delta }_{r}}^{-1}(\omega ){{A}_{\rho}}\Delta (\omega ){{\Lambda }^{-1}}(\omega ) \\
	\end{aligned} 
	\end{align}
	where $G_n(\omega)$ is the $n^{\text{th}}$ harmonic describing function for sinusoidal input with frequency of $\omega$.
	\begin{figure}[t!]
		\centering
		\includegraphics[width=\columnwidth]{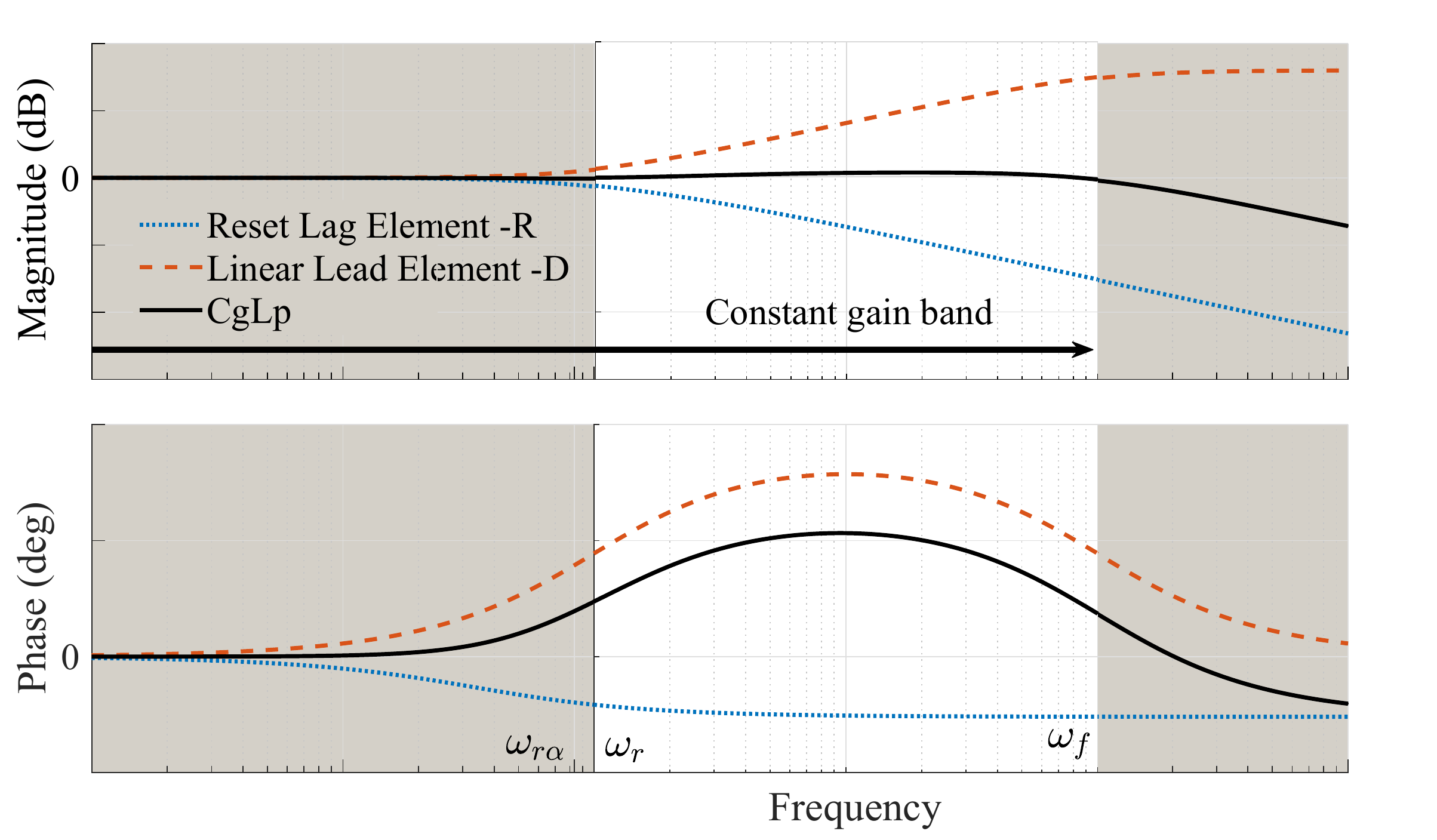}
		\caption{The concept of using combination of a reset lag and a linear lead element to form a CgLp element. The figure is adopted from~\cite{saikumar2019constant}.}
		\label{fig:cglp}
	\end{figure}
	\subsection{CgLp}
	According to~\cite{saikumar2019constant}, CgLp is a broadband phase compensation element whose first harmonic gain behaviour is constant while providing a phase lead. Originally, two architectures for CgLp are suggested using FORE or SORE, both consisting in a reset lag element in series with a linear lead filter, namely $R$ and $D$. For FORE CgLp:
	\begin{align}
	\label{eq:fore}
	&R(s)=\cancelto{A_\rho}{\frac{1}{{s}/{{{\omega }_{r\alpha }}+1}\;}},&D(s)=\frac{{s}/{{{\omega }_{r}}}\;+1}{{s}/{{{\omega }_{f}}}\;+1}
	\end{align}
	For SORE CgLp:
	\begin{equation}
	\label{eq:sore}
	\begin{aligned}
	&R(s)=\cancelto{A_\rho}{\frac{1}{({s}/{{{\omega }_{r\alpha }}{{)}^{2}}+(2s{{{\beta }}}/{{{\omega }_{r\alpha }})}\;+1}\;}}\\ \\
	&D(s)=\frac{({s}/{{{\omega }_{r}}{{)}^{2}}+(2s{{{\beta }}}/{{{\omega }_{r}})}\;+1}\;}{({s}/{{{\omega }_{f}}{{)}^{2}}+(2s{{}}/{{{\omega }_{f}})}\;+1}\;}    
	\end{aligned}
	\end{equation}
	In~(\ref{eq:fore}) and~(\ref{eq:sore}), $\omega_{r\alpha}=\omega_r/\alpha$, $\alpha$ is a tuning parameter accounting for a shift in corner frequency of the filter due to resetting action, $\beta$ is the damping coefficient and $[\omega_{r},\omega_{f}]$ is the frequency range where the CgLp will provide the required phase lead. The arrow indicates that the states of element are reset according to $A_\rho$; i.e. are multiplied by $A_\rho$ when the reset condition is met.\\
	The main idea behind the CgLp is taking the phase advantage of reset lag element over its linear counter part and use it in combination with a corresponding lead element to create broadband phase lead. Ideally, the gain of the reset lag element should be cancelled out by the gain of the corresponding linear lead element, which create a constant gain behaviour. The concept is depicted in Fig.~\ref{fig:cglp}.\\
	It can be seen that since this idea is based on DF approximation, the ideal behaviour of CgLp will not be achieved when DF is not a reliable approximation, i.e., when the higher-order harmonics are relatively large and not negligible. Nevertheless, the main idea of CgLp is not restricted to FORE and SORE by nature and can be generalized to any reset lag and linear lead filter. This paper uses fractional ones to reduce higher order harmonics in a large range of frequencies and consequently create a CgLp element that has close-to-ideal behaviour in a larger range of frequencies.
	\subsection{Second-Order Single State Reset Element (SOSRE)}
	This reset element is a special case of a SORE, in which only one integrator resets in a specific architecture. This element is presented in~\cite{karbasizadeh2020benefiting} and is used in the framework of a CgLp as the reset lag element. The state space representation and the reset matrix of the element is as follows:
	\begin{align}\nonumber
	& A=\left[ \begin{matrix}
	0 & 1   \\
	-\omega _{r\alpha }^{2} & -2{{\beta }}{{\omega }_{r\alpha }}   \\
	\end{matrix} \right],B=\left[ \begin{matrix}
	0  \\
	1  \\
	\end{matrix} \right], 
	C=\left[ \begin{matrix}
	\omega_{r\alpha } & 0  \\
	\end{matrix} \right], 
	D=\left[ 0 \right],\\
	& A_\rho=\left[ \begin{matrix}
	1 & 0   \\
	0 & \gamma   \\
	\end{matrix} \right].
	\end{align}  
	As shown in~\cite{karbasizadeh2020benefiting}, assuming a sinusoidal input at frequency of $\omega_{r\alpha}$ to a SOSRE, the steady-state output will be a sinusoidal with the same frequency and no phase shift; thus, the magnitude of higher-order harmonics at $ \omega_{r\alpha}$ is zero. 
	\subsection{Fractional order calculus and CRONE approximation of $s^\lambda$}
	Fractional order calculus developed by generalizing the integration and differentiation to non-integer order operators. Behaviour of such an element should be approximated for application in control. This paper uses CRONE approximation of ${{s}^{\lambda }},\text{ }\lambda \in {{\Re }^{-}}$, which creates fractional behaviour using real stable poles and real minimum phase zeros for this purpose. The approximation is valid in a frequency range of $[{{\omega }_{l}},{{\omega }_{h}}]$. Referring to~\cite{Oustaloup1991}, the approximation will be:
	\begin{align}
	&s^\lambda \approx C \prod_{m=1}^{N} \frac{1+\frac{s}{\omega_{z,m}}}{1+\frac{s}{\omega_{p,m}}}
	\label{eq:04:crone} \\
	&\omega_{z,m} = \omega_l \left( \frac{\omega_h}{\omega_l} \right)^{\frac{2m-1-\lambda}{2N}}
	\label{eq:04:crone:zeros} \\
	&\omega_{p,m} = \omega_l \left( \frac{\omega_h}{\omega_l} \right)^{\frac{2m-1+\lambda}{2N}}
	\label{eq:04:crone:poles}
	\end{align}
	where $N$ is number poles and zeros and for an acceptable approximation it should one unit more than the number of the decades in approximation. CRONE makes sure that the poles and zeros are placed in equal distance in logarithmic scale. $C$ is the tuning parameter for adjusting the gain of the approximation. Considering the range where approximation is valid, CRONE is actually approximating the $\left(\frac{\frac{s}{\omega_l}+1}{\frac{s}{\omega_h}+1}\right)^\lambda$. Assuming a large enough $\omega_h$, in this paper the CRONE is used to approximate $\left({\frac{s}{\omega_l}+1}\right)^\lambda$.
	\begin{figure}[t!]
		\centering
		\resizebox{\columnwidth}{!}{

			\tikzset{every picture/.style={line width=0.75pt}} %set default line width to 0.75pt        
			
			\begin{tikzpicture}[x=0.75pt,y=0.75pt,yscale=-1,xscale=1]
			%uncomment if require: \path (0,249.6666717529297); %set diagram left start at 0, and has height of 249.6666717529297
			
			%Flowchart: Alternative Process [id:dp7007021524924317] 
			\draw  [fill={rgb, 255:red, 0; green, 0; blue, 0 }  ,fill opacity=0.05 ][dash pattern={on 4.5pt off 4.5pt}] (21.81,81.3) .. controls (21.81,64.6) and (35.35,51.06) .. (52.05,51.06) -- (358.05,51.06) .. controls (374.75,51.06) and (388.29,64.6) .. (388.29,81.3) -- (388.29,193.61) .. controls (388.29,210.31) and (374.75,223.85) .. (358.05,223.85) -- (52.05,223.85) .. controls (35.35,223.85) and (21.81,210.31) .. (21.81,193.61) -- cycle ;
			%Shape: Rectangle [id:dp8144323491127636] 
			\draw   (120.5,88) -- (167.5,88) -- (167.5,132.22) -- (120.5,132.22) -- cycle ;
			%Straight Lines [id:da39247231885225653] 
			\draw    (111,144) -- (174.99,73.23) ;
			\draw [shift={(177,71)}, rotate = 492.12] [fill={rgb, 255:red, 0; green, 0; blue, 0 }  ][line width=0.08]  [draw opacity=0] (10.72,-5.15) -- (0,0) -- (10.72,5.15) -- (7.12,0) -- cycle    ;
			
			%Flowchart: Summing Junction [id:dp8580452502150646] 
			\draw   (70.83,107.67) .. controls (70.83,103.25) and (74.64,99.67) .. (79.33,99.67) .. controls (84.03,99.67) and (87.83,103.25) .. (87.83,107.67) .. controls (87.83,112.08) and (84.03,115.67) .. (79.33,115.67) .. controls (74.64,115.67) and (70.83,112.08) .. (70.83,107.67) -- cycle ; \draw   (73.32,102.01) -- (85.34,113.32) ; \draw   (85.34,102.01) -- (73.32,113.32) ;
			%Straight Lines [id:da45878652535418984] 
			\draw    (11.5,107.33) -- (42.49,107.49) -- (67.83,107.65) ;
			\draw [shift={(70.83,107.67)}, rotate = 180.36] [fill={rgb, 255:red, 0; green, 0; blue, 0 }  ][line width=0.08]  [draw opacity=0] (8.93,-4.29) -- (0,0) -- (8.93,4.29) -- cycle    ;
			%Straight Lines [id:da07056757658428237] 
			\draw    (87.83,107.67) -- (88.49,107.82) -- (120.67,108.22) ;
			%Shape: Rectangle [id:dp710693576641573] 
			\draw   (214.5,88.33) -- (281.53,88.33) -- (281.53,132.22) -- (214.5,132.22) -- cycle ;
			%Straight Lines [id:da8549134122358593] 
			\draw    (167.83,108.33) -- (168.49,108.49) -- (214.67,108.89) ;
			%Flowchart: Summing Junction [id:dp1273203267260099] 
			\draw   (70.83,165.67) .. controls (70.83,161.25) and (74.64,157.67) .. (79.33,157.67) .. controls (84.03,157.67) and (87.83,161.25) .. (87.83,165.67) .. controls (87.83,170.08) and (84.03,173.67) .. (79.33,173.67) .. controls (74.64,173.67) and (70.83,170.08) .. (70.83,165.67) -- cycle ; \draw   (73.32,160.01) -- (85.34,171.32) ; \draw   (85.34,160.01) -- (73.32,171.32) ;
			%Straight Lines [id:da0003761026806807788] 
			\draw    (79.33,118.67) -- (79.33,157.67) ;
			\draw [shift={(79.33,115.67)}, rotate = 90] [fill={rgb, 255:red, 0; green, 0; blue, 0 }  ][line width=0.08]  [draw opacity=0] (8.93,-4.29) -- (0,0) -- (8.93,4.29) -- cycle    ;
			%Straight Lines [id:da9656233599884168] 
			\draw    (90.83,165.71) -- (112,166) ;
			\draw [shift={(87.83,165.67)}, rotate = 0.79] [fill={rgb, 255:red, 0; green, 0; blue, 0 }  ][line width=0.08]  [draw opacity=0] (8.93,-4.29) -- (0,0) -- (8.93,4.29) -- cycle    ;
			%Flowchart: Extract [id:dp08329342420891939] 
			\draw   (112,166) -- (180,146.81) -- (180,185.19) -- cycle ;
			%Straight Lines [id:da6796548737257648] 
			\draw    (191.58,108.69) -- (191.33,166.22) ;
			%Straight Lines [id:da9844774327530221] 
			\draw    (180,165.89) -- (191.33,166.22) ;
			%Straight Lines [id:da9913051930442374] 
			\draw    (281.19,108.87) -- (311.29,108.8) ;
			%Flowchart: Extract [id:dp5022270089564815] 
			\draw   (379.67,108.67) -- (311.67,89.47) -- (311.67,127.86) -- cycle ;
			%Flowchart: Extract [id:dp7675121692815141] 
			\draw   (184,198) -- (252,178.81) -- (252,217.19) -- cycle ;
			%Straight Lines [id:da7695942530087705] 
			\draw    (296.08,108.69) -- (295.33,198.22) ;
			%Straight Lines [id:da7675842968745559] 
			\draw    (252,198.22) -- (295.33,198.22) ;
			%Straight Lines [id:da005369306010184038] 
			\draw    (79.33,198.22) -- (184,198) ;
			%Straight Lines [id:da42818530288498513] 
			\draw    (79.33,176.67) -- (79.33,198.22) ;
			\draw [shift={(79.33,173.67)}, rotate = 90] [fill={rgb, 255:red, 0; green, 0; blue, 0 }  ][line width=0.08]  [draw opacity=0] (8.93,-4.29) -- (0,0) -- (8.93,4.29) -- cycle    ;
			%Straight Lines [id:da3926532307063242] 
			\draw  [dash pattern={on 4.5pt off 4.5pt}]  (42.49,107.49) -- (42.67,127.56) ;
			%Straight Lines [id:da08935274173330687] 
			\draw  [dash pattern={on 4.5pt off 4.5pt}]  (42.67,127.56) -- (120.67,126.89) ;
			%Shape: Rectangle [id:dp3604544615913692] 
			\draw   (415.5,86.33) -- (462.5,86.33) -- (462.5,130.22) -- (415.5,130.22) -- cycle ;
			%Straight Lines [id:da42069424743356953] 
			\draw    (379.67,108.67) -- (416.5,109.33) ;
			%Straight Lines [id:da9355922753288886] 
			\draw    (461.67,107.67) -- (490.5,107.33) ;
			
			\begin{large}
			% Text Node
			\draw (43,94) node    {$e( t)$};
			% Text Node
			\draw (196,97) node    {$x_{2}$};
			% Text Node
			\draw (180,64) node    {\Large$\gamma $};
			% Text Node
			\draw (248,110.28) node    {$\left(\frac{s}{\omega _{l}} +1\right)^{\lambda }$};
			% Text Node
			\draw (300,97.67) node    {$x_{1}$};
			% Text Node
			\draw (158.67,167.67) node    {$2\beta \omega _{r\alpha }$};
			% Text Node
			\draw (333,108) node    {$\omega ^{2}_{r\alpha }$};
			% Text Node
			\draw (234,197.33) node    {$\omega ^{2}_{r\alpha }$};
			% Text Node
			\draw (68,94.67) node    {$+$};
			% Text Node
			\draw (89.33,120.67) node    {$-$};
			% Text Node
			\draw (90,153) node    {$+$};
			% Text Node
			\draw (68.67,174.67) node    {$+$};
			% Text Node
			\draw (439,108.28) node    {$D( s)$};
			% Text Node
			\draw (482,93) node    {$u( t)$};
			% Text Node
			\draw (195,27) node   [align=center] {Fractional-Order Single\\State Reset Element};
			% Text Node
			\draw (144,107.5) node    {\LARGE$\frac{1}{s}$};
			\end{large}
			
			\end{tikzpicture}
			
		}
		\caption{Block diagram of the a FOSRE CgLp. $\lambda \in (0~\text{  }~{-1}]$.}
		\label{fig:fosre_block}
	\end{figure}
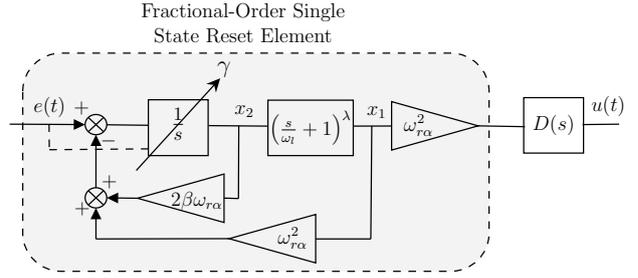
	\section{Fractional-Order Single State Reset Element (FOSRE)}
	This sections introduces a new structure for reset elements in the framework of CgLp and discusses the architecture, frequency response and its superiority over FORE and SORE in the framework of CgLp. 
	\subsection{Architecture}
	The architecture of the FOSRE is similar to that of the SORE with the difference being that the second linear integrator is replaced with a fractional one, and only first integrator, which is a linear one is reset. Figure~\ref{fig:fosre_block} shows the block diagram of the element. The following defines the FOSRE (the reset lag element) and its corresponding lead element to form a CgLp. 
	\begin{equation}
	\label{eq:fosre}
	\begin{aligned}
	&FOSRE(s)=\cancelto{A_\rho}{\frac{1}{{{\left( s/{{\omega }_{l}}+1 \right)}^{-\lambda }}\left( s/\omega _{r\alpha }^{2}+2\beta /{{\omega }_{r\alpha }} \right)+1}}\\ \\
	&D(s)=\frac{{{\left( s/{{\omega }_{l}}+1 \right)}^{-\lambda }}\left( s/\omega _{r }^{2}+2\beta /{{\omega }_{r }} \right)+1}{({s}/{{{\omega }_{f}}{{)}^{2}}+(2s{{}}/{{{\omega }_{f}})}\;+1}\;}    
	\end{aligned}
	\end{equation}
	A matching state space representation of FOSRE with the architecture of Fig.~\ref{fig:fosre_block} is as follows:
	\begin{align}
	\label{eq:fosre_ss}\nonumber
	& A=\left[ \begin{matrix}
	-2\beta {{\omega }_{r\alpha }} & {{0}_{1\times N}}  \\
	\overline{{{B}}}_{N\times 1} & \overline{{{A}}}_{N\times N}  \\
	\end{matrix} \right]-\omega _{r\alpha }^{2}\left[ \begin{matrix}
	1  \\
	{{0}_{N\times 1}}  \\
	\end{matrix} \right]\times \left[ \begin{matrix}
	\overline{{{D}}}_{1\times 1} & \overline{{{C}}}_{1\times N}  \\
	\end{matrix} \right], \\ 
	& B=\left[ \begin{matrix}
	1  \\
	{{0}_{N\times 1}}  \\
	\end{matrix} \right],\quad C=\omega_{r \alpha}^2\left[ \begin{matrix}
	\overline{{{D}}}_{1\times 1} & \overline{{{C}}}_{1\times N}  \\
	\end{matrix} \right], \quad D=0\\ \nonumber
	& {{A}_{\rho }}=\text{diag}(\gamma ,\underbrace{1,...,1}_{N})
	\end{align}
	where $\overline{A}$, $\overline{B}$, $\overline{C}$ and $\overline{D}$ are state space matrices of the CRONE approximation of $\left({\frac{s}{\omega_l}+1}\right)^\lambda$.
	\subsection{Linear behaviour of FOSRE at a certain frequency}
	The architecture of FOSRE, along with the non-identical reset of its states, creates a peculiar phenomenon which can be used to the benefit of the performance of the system.  
	\begin{lemma}
		\label{lemma1}    
		Reset control system Eq.~(\ref{eq:reset}) in open-loop has a globally asymptotically stable $2\pi / \omega$-periodic solution under sinusoid input with arbitrary frequency, $\omega>0$ if and only if
		\begin{align}
		\left| \lambda \left(A_\rho e^{A \delta}\right)\right|<1 \quad \forall \delta \in {{\Re }^{+}}
		\end{align}
	where $\lambda\left(.\right)$ stands for eigenvalues~\cite{guo2009frequency}.
		\end{lemma}
	\begin{figure}[t!]
		\centering
		\includegraphics[width=\columnwidth]{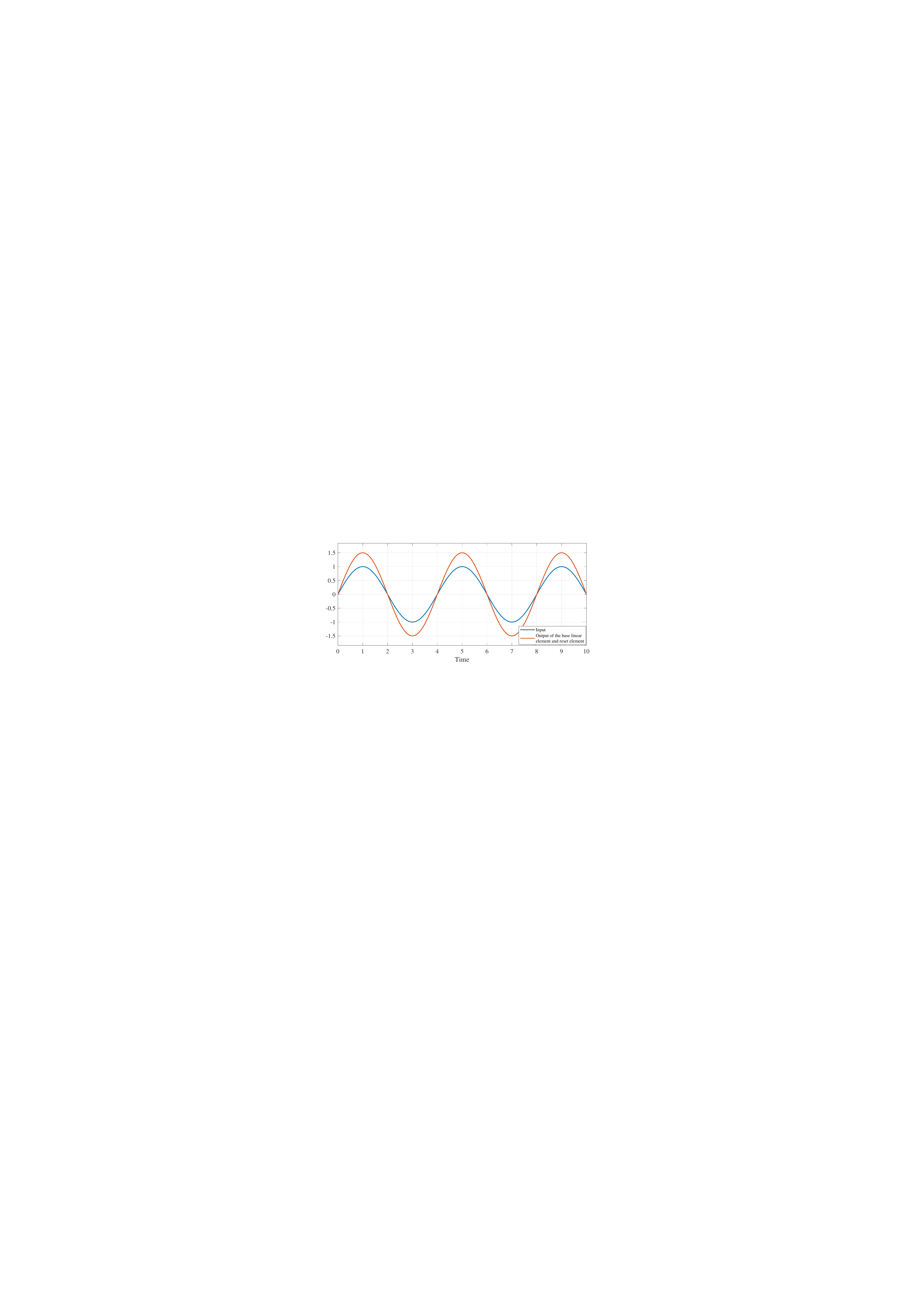}
		\caption{Assuming that the output of the base linear element for a reset element has no phase shift with respect to its input, the output of the reset element itself will match the base linear element output at steady state.}
		\label{fig:no_phase_shift}
	\end{figure}
	\begin{remark}
		\label{remark1}    
		Let us define
		\begin{equation}
		\psi(\omega) :=\angle \frac{X_2(j\omega)}{E(j\omega)}\quad \text{for } A_\rho=I.
		\end{equation} 
		Assuming a sinusoidal input, $\sin(\omega_{lb} t)$, to a reset element, the reset action will be of no effect in steady state response, and thus the reset element can be regarded as a linear system in terms of steady state response at that certain frequency if:
		\begin{equation}
		\psi(\omega_{lb}) =0.
		\end{equation} 
	\end{remark} 
	The proof of this is trivial, since the reset element under such circumstances will reset its output, when its output is at zero, resulting in no change from the resetting action. Figure~\ref{fig:no_phase_shift} shows an example of this situation, where steady state output of the base linear element and the reset element itself will be the same. At such circumstance, the reset element can be regarded as a linear element at that certain frequency in terms of steady state output.\\
	\begin{remark}
		\label{remark2}    
		Assuming a sinusoidal input, $\sin(\omega_{lb} t)$, to a reset element where $\psi(\omega_{lb})=0$, the higher-order harmonics will be zero.
	\end{remark}
	\begin{figure}[t!]
		\centering
		\includegraphics[width=\columnwidth]{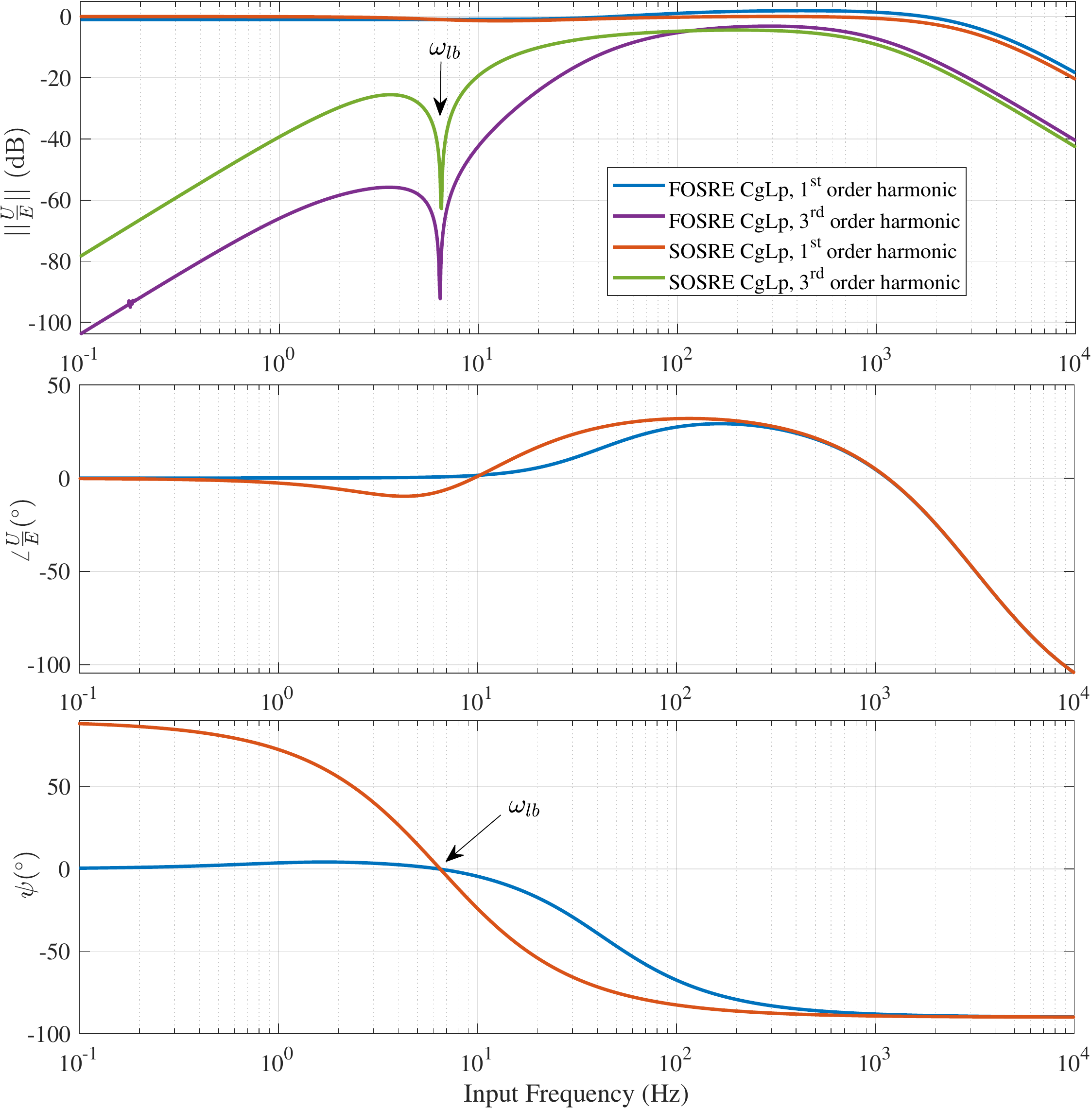}
		\caption{HOSIDF comparison of a SOSRE and a FOSRE CgLp along with corresponding $\psi$ which is $\angle \frac{X_2}{E}$ of the base linear system. For FOSRE CgLp, $\omega_{r\alpha }=3.18\text{ Hz}, \beta=1, \alpha=0.94, \omega_{l}=0.8\text{ Hz}$, $\lambda=-0.1$ and $\gamma=0.2$. For SOSRE CgLp, $\omega_{r\alpha }=6.5\text{ Hz}, \beta=1, \alpha=1.12$ and $\gamma=0.2$.}
		\label{fig:FOSRE_HOSIDF_bpp}
	\end{figure}
	Since the reset action has no effect at steady-state, the steady-state output is sinusoidal. Such an output can be completely described by first harmonic of Fourier series and all higher-order harmoincs are zero.\\
	For the case of a FOSRE, if $e(t)=\sin(\omega_{lb} t)$, the reset action of the first integrator will be of no effect if:
	\begin{align}\nonumber
	%&\angle \frac{{{X}_{2}}(j\omega_{lb} )}{E(j\omega_{lb} )}=0 \Rightarrow\\
	&\psi(\omega_{lb} )=0 \Rightarrow\\ \nonumber
	&\angle \frac{1}{j\omega_{lb} /\omega _{r\alpha }^{2}+2\beta /{{\omega }_{r\alpha }}+{{\left( j\omega_{lb} /{{\omega }_{l}}+1 \right)}^{\lambda }}}=0 \Rightarrow\\ 
	&{{\omega }_{lb}}=-\omega _{r\alpha}^{2}{{\left( \frac{\omega _{lb}^{2}}{\omega _{l}^{2}}+1 \right)}^{\frac{\lambda }{2}}}\sin \left( \lambda {{\tan }^{-1}}\left( \frac{{{\omega }_{lb}}}{{{\omega }_{l}}} \right) \right).
	\label{eq:w_lb}
	\end{align}
	Thus, FOSRE exhibits a linear behaviour at a frequency, $\omega_{lb}$, which depends on $\omega_{r \alpha}$, $\beta$, $\omega_l$ and $\lambda$.
	\subsection{HOSIDF of FOSRE CgLp}
	Reset elements are nonlinear elements because of the discontinuity in their state values and output. This discontinuity or in other term, jumps, creates higher-order terms in Fourier decomposition of the output. These jumps also create large peaks in controller's output which is a known characteristic for reset elements.\\
	%HOSIDF method takes the higher-order harmonics into account on top of the first-order one. Studying higher-order order harmonics is important as they can play a negative role in overall closed-loop performance.\\ % It can be concluded that the smaller higher order harmonics are, the closer the real controller is to its design based on DF.\\
	There are many situations where the system's behaviour  is not predictable based on DF. As an example, for the mass-spring-damper systems with a resonance peak at $\omega_n$, the $3^\text{rd}$ HOSIDF of the open loop system has a peak at  $\omega_n/3$ and the $5^\text{th}$ HOSIDF has a peak at  $\omega_n/5$ and so on. If the resonance peak of the system is large enough, higher-order harmonics will probably dominate the first one at  $\omega_n/3$,  $\omega_n/5$, ... . (See~\cite{karbasizadeh2020benefiting, kars2018HOSIDF}). Therefore, it can be concluded that the smaller higher-order harmonics are, the closer the system is to what it is designed for.\\
	\begin{figure}[t!]
		\centering
		\includegraphics[width=\columnwidth]{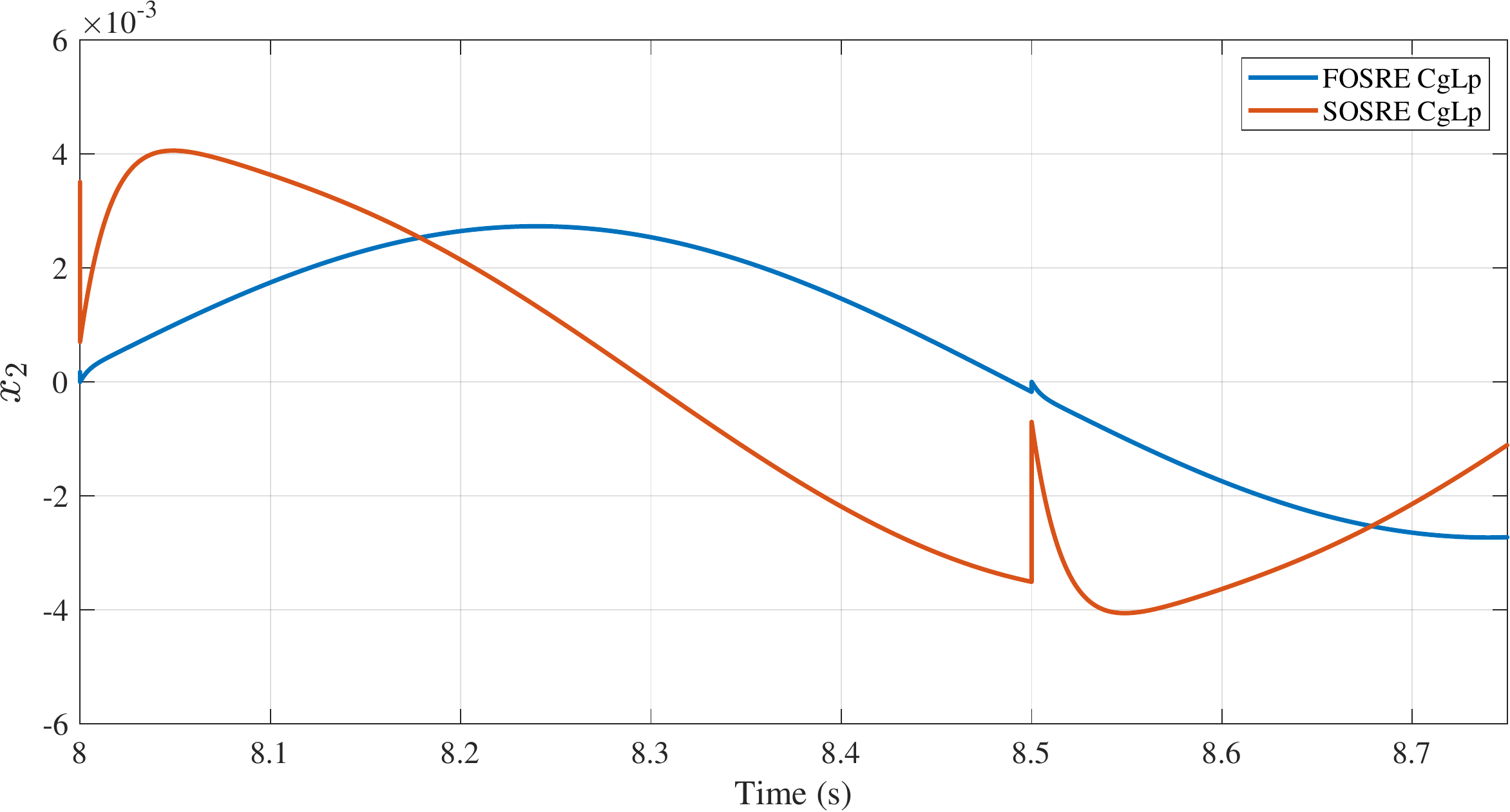}
		\caption{Comparing $x_2$ of the FOSRE CgLp and SOSRE CgLp for an input of $\sin (2\pi t)$.}
		\label{fig:FOSRE_time_compare}
	\end{figure} 
	The main benefit of using FOSRE is that tuning its parameters, one can reduce higher-order harmonics at a range of frequency where nonlinearity not only does not have a clear benefit but also deteriorates the tracking precision of the system.\\
	In~\cite{karbasizadeh2020benefiting}, it was shown that eliminating higher-order harmonics using the concept of Remark~\ref{remark1} at one frequency in a SOSRE CgLp element, results in improvement of steady-state tracking precision. While FOSRE enjoys the same benefit, HOSIDF analysis in this section shows that structure of FOSRE allows reduction of the higher-order harmonics at a wider range of frequencies.\\
	In FOSRE, the concept of Remark~\ref{remark1} can be generalised. According to Remark~\ref{remark1}, when $\psi$ is zero, the higher-order harmonics will be zero. On top of that, it can be seen empirically that closer the  $\psi$ is to zero, smaller the gain of the higher-order harmonics is. Figure~\ref{fig:FOSRE_HOSIDF_bpp} shows this relation by comparing HOSIDF of FOSRE and a SOSRE CgLp along with their $\psi$ plot. All the even harmonics are zero for reset elements and Fig.~\ref{fig:FOSRE_HOSIDF_bpp} only depicts the $1^\text{st}$ and the $3^\text{rd}$ harmonic for the sake of the clarity of the figure since all the other odd harmonics will follow the same trend as the $3^\text{rd}$ one and are descending with respect to their order.\\
	Three main conclusions can be drawn from this figure. 
	\begin{itemize}
		\item Both of the $\psi$ plots cross zero between 6 and 7~Hz and correspondingly higher-order harmonics will be zero at $\omega_{lb}$ which validates Remark~\ref{remark1} and~\ref{remark2}. 
		\item For the range of 0.1 till 500~Hz, $\psi$ of FOSRE CgLp is closer to zero than that of SOSRE CgLp, correspondingly its magnitude of higher-order harmonics is smaller than SOSRE CgLp. 
		\item As the $\psi$ approaches more negative values, the phase advantage of CgLp elements increases. Hence, comparing two CgLp elements, especially between 10 and 100 Hz, the closer value of $\psi$ to zero results in less phase advantage.
	\end{itemize}
	\begin{figure}[t!]
		\centering
		\includegraphics[width=\columnwidth]{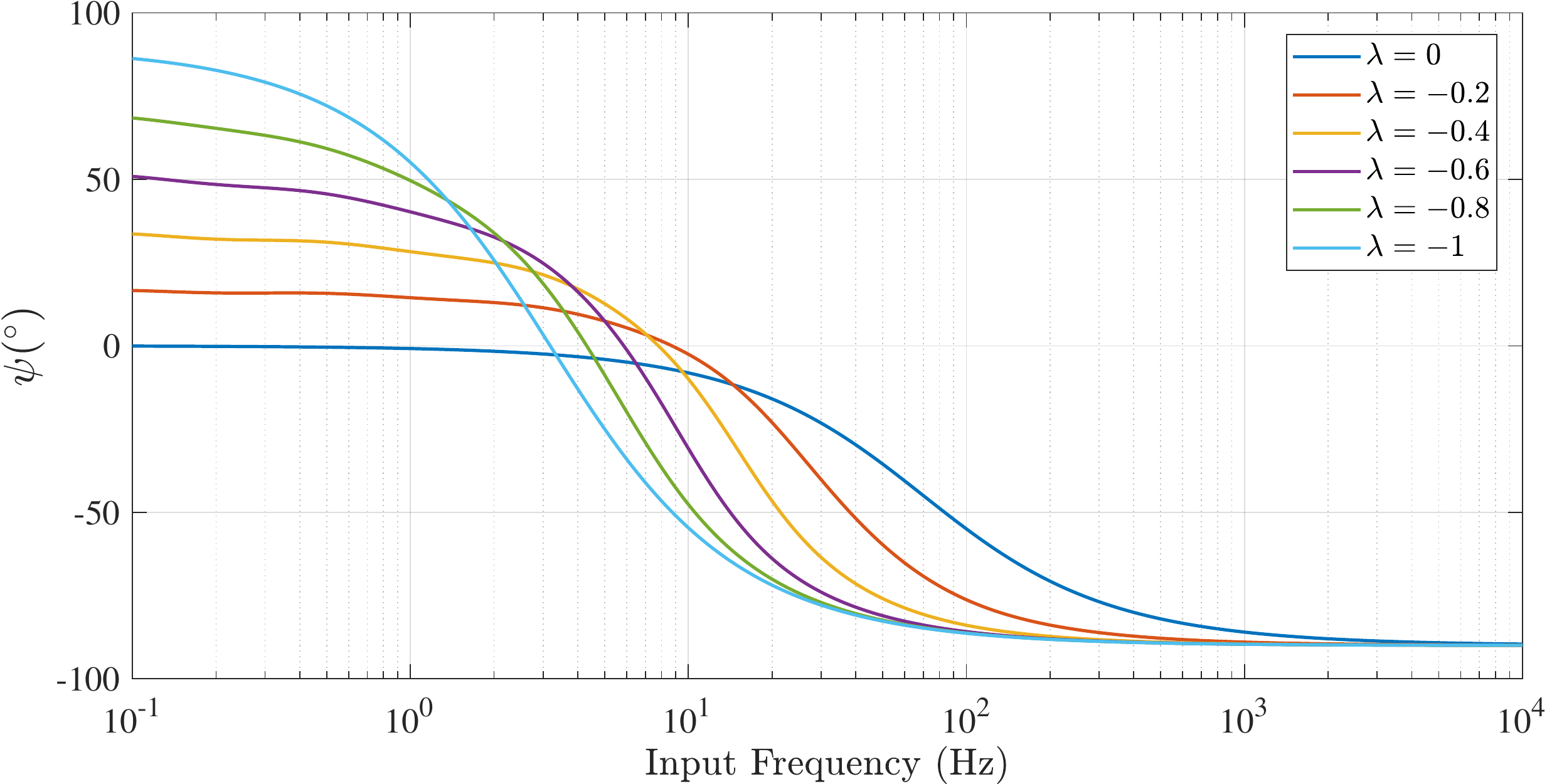}
		\caption{The effect of $\lambda$ on $\psi$ in FOSRE. $\omega_{r\alpha }=3.18\text{ Hz}, \beta=1$ and $\omega_{l}=0.001\text{ Hz}$.}
		\label{fig:lambda}
	\end{figure}
	The same line of reasoning has been tested and is valid comparing any two FOSRE CgLp elements. The relation between $\psi$ and higher-order harmonics can be justified by considering the fact that closer the $\psi$ to zero is, smaller the jumps of the reset element will be. Furthermore, the biggest jump and thus, the largest higher-order harmonics will happen when $\psi=\pm 90^\circ$. This concept is shown in Fig.~\ref{fig:FOSRE_time_compare}, by comparing the time response of $x_2$ in FOSRE CgLp and SOSRE CgLp to a sinusoidal input of 1~Hz. It is readily obvious that due to the smaller value of $\psi$ in FOSRE CgLp at this frequency, the jumps are smaller and thus justifies the smaller magnitude of higher-order harmonics.\\
	According to aforementioned discussion, $\psi$ plot contains important information about higher-order harmonics and phase advantage created by CgLp elements, and hence can be used to tune the FOSRE parameters so that FOSRE CgLp element has a closer-to-ideal behaviour.\\
	The following section will discuss the effect of FOSRE parameters on $\psi$ plot and thus higher-order harmonics.
	\begin{figure}[t!]
		\centering
		\includegraphics[width=\columnwidth]{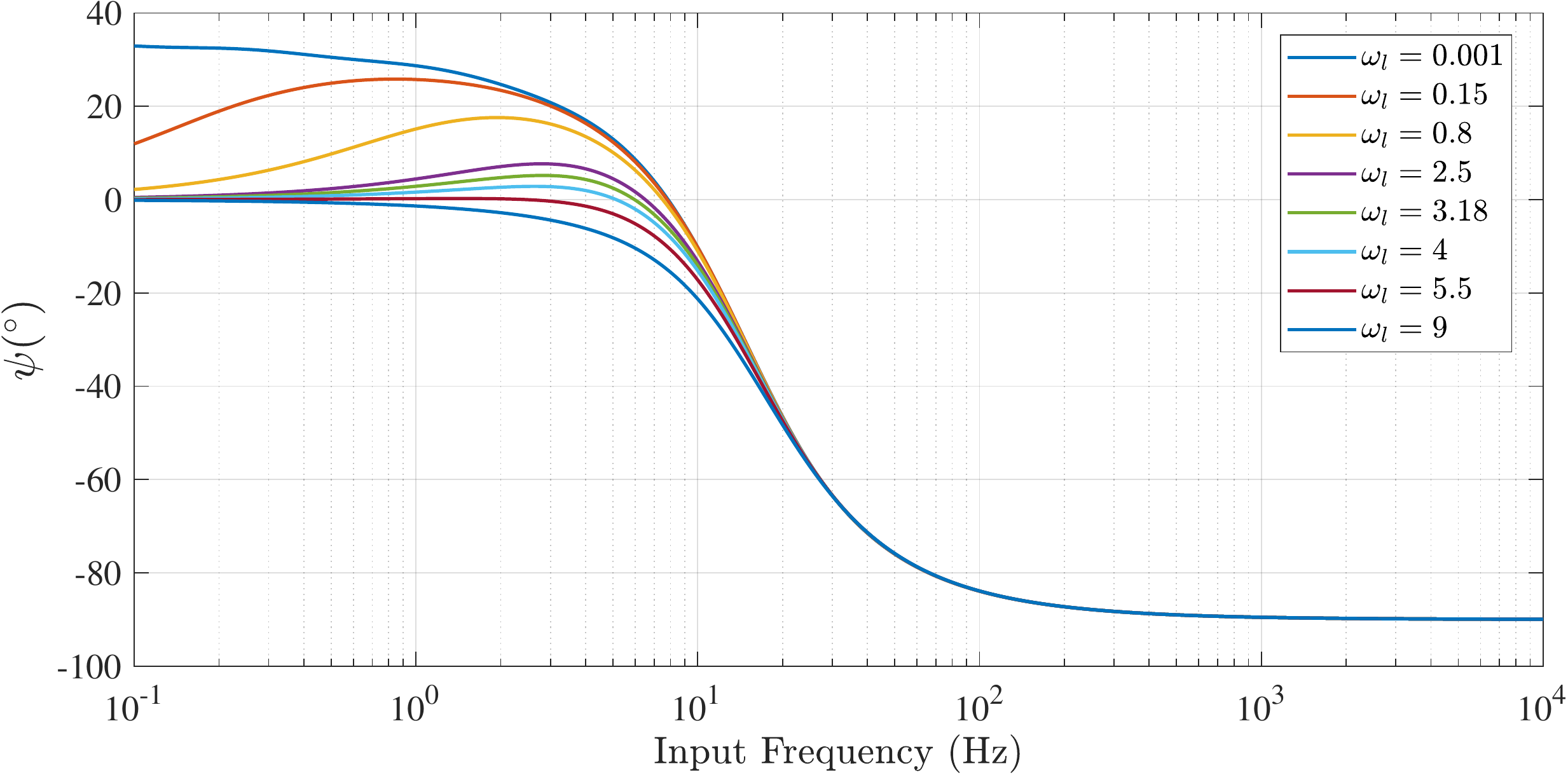}
		\caption{The effect of $\omega_l$ on $\psi$ in FOSRE. $\omega_{r\alpha }=3.18\text{ Hz}, \beta=1$ and $\lambda=-0.4$.}
		\label{fig:w_l}
	\end{figure} 
	\section{Suppressing higher-order harmonics at low frequencies}
	\label{sec:parameters}
	The main advantage of FOSRE with respect to SOSRE is that the nonlinearity effects, i.e., the higher-order harmonics, can be suppressed at low frequencies. This can be done by manipulating the $\psi$ values at lower frequencies. This is made possible for FOSRE by additional two parameters, namely, $\lambda$ and $\omega_{l}$. In the following, the effects of these two parameters on the $\psi$ plot are discussed.\\ 
	The effect of $\lambda$ on higher-order harmonics and phase advantage created by a FOSRE can be depicted by plotting $\psi$ versus input frequency for different values of $\lambda$. See Fig.~\ref{fig:lambda}. As $\lambda$ approaches -1, $\psi$ deviates more from zero in lower frequencies which indicates larger higher-order harmonics at low frequencies and thus deterioration of tracking precision. Nevertheless, $\psi$ will approach $-90^\circ$ faster, in turn, phase advantage of CgLp element will be available at a wider range of frequency. For example, designing a controller for 100 Hz bandwidth, a FOSRE depicted in Fig.~\ref{fig:lambda} with $\lambda=-0.2$ will have less phase margin comparing to the one with $\lambda=-0.8$. Notice that $\omega_{lb}$ varies with $\lambda$ and for $\lambda=0$, $\omega_{lb}=0$ which means FOSRE will not show linear behaviour for such a configuration.\\
	Changing $\lambda$, suppressing higher-order harmonics comes at the cost of losing phase advantage. However, tuning $\omega_{l}$, one can circumvent this limitation. Figure~\ref{fig:w_l} depicts $\psi$ plot for different values of $\omega_{l}$ when $\lambda=-0.4$. Increasing $\omega_{l}$ to a certain point, $\psi$ gets closer to zero for frequencies below $\omega_{lb}$, while it will not cause loss of phase advantage in the crossover frequency region. It should be noticed that $\omega_{lb}$ according to~(\ref{eq:w_lb}), depends on $\omega_{l}$.\\
	The ideal tracking performance of a CgLp will happen at $\omega_{lb}$ where higher-order harmonics are zero, and hence, the tracking error can be accurately calculated using DF. If tracking at a certain frequency is important for a system, one may consider designing $\omega_{lb}$ to match that frequency. Otherwise, as suggested in~\cite{karbasizadeh2020benefiting}, $\omega_{lb}$ can be designed to match and cancel out the peak of the $3^\text{rd}$ order harmonic.\\ 
	\begin{figure}[t!]
		\centering
		\includegraphics[width=\columnwidth]{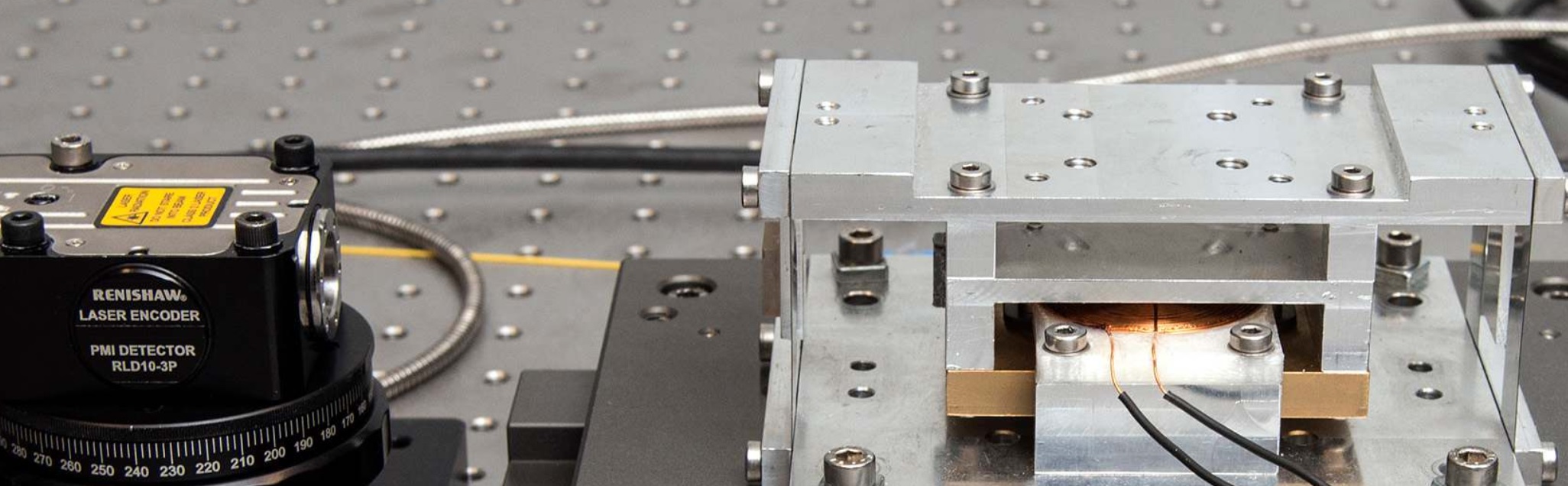}
		\caption{The stage whose transfer function is used for simulation.}
		\label{fig:setup}
	\end{figure} 
	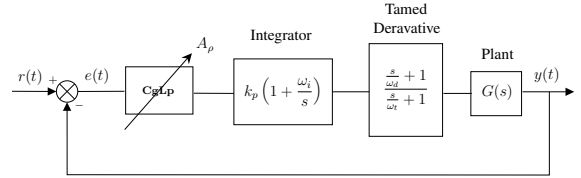
\begin{figure}[t!]

		\centering
		\scalebox{0.9}[0.9]{
			\resizebox{\columnwidth}{!}{
				
				\tikzset{every picture/.style={line width=0.75pt}} %set default line width to 0.75pt        
				
				\begin{tikzpicture}[x=0.75pt,y=0.75pt,yscale=-1,xscale=1]
				%uncomment if require: \path (0,203); %set diagram left start at 0, and has height of 203
				
				%Shape: Rectangle [id:dp9250873333024565] 
				\draw  [line width=0.75]  (365.21,47.25) -- (440.5,47.25) -- (440.5,143) -- (365.21,143) -- cycle ;
				
				%Shape: Rectangle [id:dp22823692677968466] 
				\draw  [line width=0.75]  (115.5,71.17) -- (184.5,71.17) -- (184.5,120) -- (115.5,120) -- cycle ;
				%Straight Lines [id:da8451766911766478] 
				\draw [line width=0.75]    (114.5,136.08) -- (181.53,59.34) ;
				\draw [shift={(183.5,57.08)}, rotate = 491.13] [fill={rgb, 255:red, 0; green, 0; blue, 0 }  ][line width=0.08]  [draw opacity=0] (10.72,-5.15) -- (0,0) -- (10.72,5.15) -- (7.12,0) -- cycle    ;
				
				%Shape: Rectangle [id:dp3993985167748695] 
				\draw  [line width=0.75]  (226.5,63) -- (327.5,63) -- (327.5,131) -- (226.5,131) -- cycle ;
				
				%Straight Lines [id:da7731408703992304] 
				\draw [line width=0.75]    (66.42,96.67) -- (114.22,96.45) ;
				%Straight Lines [id:da36461852787237525] 
				\draw [line width=0.75]    (550,97) -- (550,183) -- (55.5,182.67) -- (55.46,110.73) ;
				\draw [shift={(55.46,107.73)}, rotate = 449.97] [fill={rgb, 255:red, 0; green, 0; blue, 0 }  ][line width=0.08]  [draw opacity=0] (8.93,-4.29) -- (0,0) -- (8.93,4.29) -- cycle    ;
				%Straight Lines [id:da1393811703303558] 
				\draw [line width=0.75]    (-1,96.45) -- (41.5,96.65) ;
				\draw [shift={(44.5,96.67)}, rotate = 180.27] [fill={rgb, 255:red, 0; green, 0; blue, 0 }  ][line width=0.08]  [draw opacity=0] (8.93,-4.29) -- (0,0) -- (8.93,4.29) -- cycle    ;
				%Straight Lines [id:da4183844429842236] 
				\draw [line width=0.75]    (522.5,97) -- (572.5,97) ;
				\draw [shift={(575.5,97)}, rotate = 180] [fill={rgb, 255:red, 0; green, 0; blue, 0 }  ][line width=0.08]  [draw opacity=0] (8.93,-4.29) -- (0,0) -- (8.93,4.29) -- cycle    ;
				%Straight Lines [id:da6169379750986348] 
				\draw [line width=0.75]    (185.42,98.45) -- (226.5,98.67) ;
				%Straight Lines [id:da8911932794469448] 
				\draw [line width=0.75]    (327.5,97) -- (365.5,97) ;
				%Shape: Rectangle [id:dp780701248947808] 
				\draw  [line width=0.75]  (469.5,76.17) -- (521.5,76.17) -- (521.5,121.67) -- (469.5,121.67) -- cycle ;
				%Straight Lines [id:da03234018932258742] 
				\draw [line width=0.75]    (441.5,98.67) -- (470.5,98.67) ;
				%Flowchart: Summing Junction [id:dp09605751187284017] 
				\draw   (44.5,96.67) .. controls (44.5,90.56) and (49.41,85.61) .. (55.46,85.61) .. controls (61.52,85.61) and (66.42,90.56) .. (66.42,96.67) .. controls (66.42,102.77) and (61.52,107.73) .. (55.46,107.73) .. controls (49.41,107.73) and (44.5,102.77) .. (44.5,96.67) -- cycle ; \draw   (47.71,88.85) -- (63.21,104.49) ; \draw   (63.21,88.85) -- (47.71,104.49) ;
				
				% Text Node
				\draw (17.84,81.76) node  [font=\large]  {$r( t)$};
				% Text Node
				\draw (547.62,79.76) node  [font=\large]  {$y( t)$};
				% Text Node
				\draw (86.35,80.76) node  [font=\large]  {$e( t)$};
				% Text Node
				\draw (496.57,97.58) node  [font=\large]  {$G( s)$};
				% Text Node
				\draw (497,58) node  [font=\normalsize] [align=left] {{\fontfamily{ptm}\selectfont {\large Plant}}};
				% Text Node
				\draw (41,85) node    {$+$};
				% Text Node
				\draw (68,111) node    {$-$};
				% Text Node
				\draw (402.85,95.13) node  [font=\large]  {$\dfrac{\frac{s}{\omega _{d}} +1}{\frac{s}{\omega _{t}} +1}$};
				% Text Node
				\draw (404,21) node  [font=\large] [align=left] {{\fontfamily{ptm}\selectfont \textbf{ \ \ }Tamed \ }\\{\fontfamily{ptm}\selectfont Deravative}};
				% Text Node
				\draw (197,48.91) node  [font=\large]  {$A_{\rho }$};
				% Text Node
				\draw (149,96.58) node   [align=left] {\textbf{CgLp}};
				% Text Node
				\draw (277,97) node  [font=\large]  {$k_{p}\left( 1+\dfrac{\omega _{i}}{s}\right)$};
				% Text Node
				\draw (275,40) node  [font=\normalsize] [align=left] {{\fontfamily{ptm}\selectfont {\large Integrator}}};

				\end{tikzpicture}
				
			}
		}
		\caption{Designed control architecture to compare the performance of two sets of controllers.} 
		\label{fig:closed-loop}
	\end{figure} 
	\begin{table}[t!]
		\centering
		\caption{Parameters of the designed controllers. All frequencies are in Hz.}
		\label{tab:params}
		\resizebox{\columnwidth}{!}{%
			\begin{tabular}{@{}llllllllll@{}}
				\toprule
				Controller & $\omega_i$ & $\omega_d$ & $\omega_t$ & $\omega_f$ & $\omega_{r\alpha}$ & $\beta$ & $\gamma$ & $\lambda$ & $\omega_l$ \\ \midrule
				PID        & 15         & 32         & 705        & 1500       & N/A                & N/A     & N/A      & N/A       & N/A        \\
				SOSRE No. 1 & 15         & 100         & 225        & 1500       & 2                  & 1       & 0.2      & N/A       & N/A        \\
				SOSRE No. 2 & 15         & 100         & 225        & 1500       & 0.8                & 1       & 0.2      & N/A       & N/A        \\
				FOSRE No. 1 & 15         & 100         & 225        & 1500       & 2                  & 1       & 0.2      & -0.4      & 2.5        \\
				FOSRE No. 2 & 15         & 100         & 225        & 1500       & 1.2                & 1       & 0.2      & -0.4      & 1.3        \\ \bottomrule
			\end{tabular}%
		}
	\end{table}
	\begin{figure*}[t!]
		\begin{subfigure}{.5\textwidth}
			\centering
			% include first image
			\includegraphics[width=0.95\linewidth]{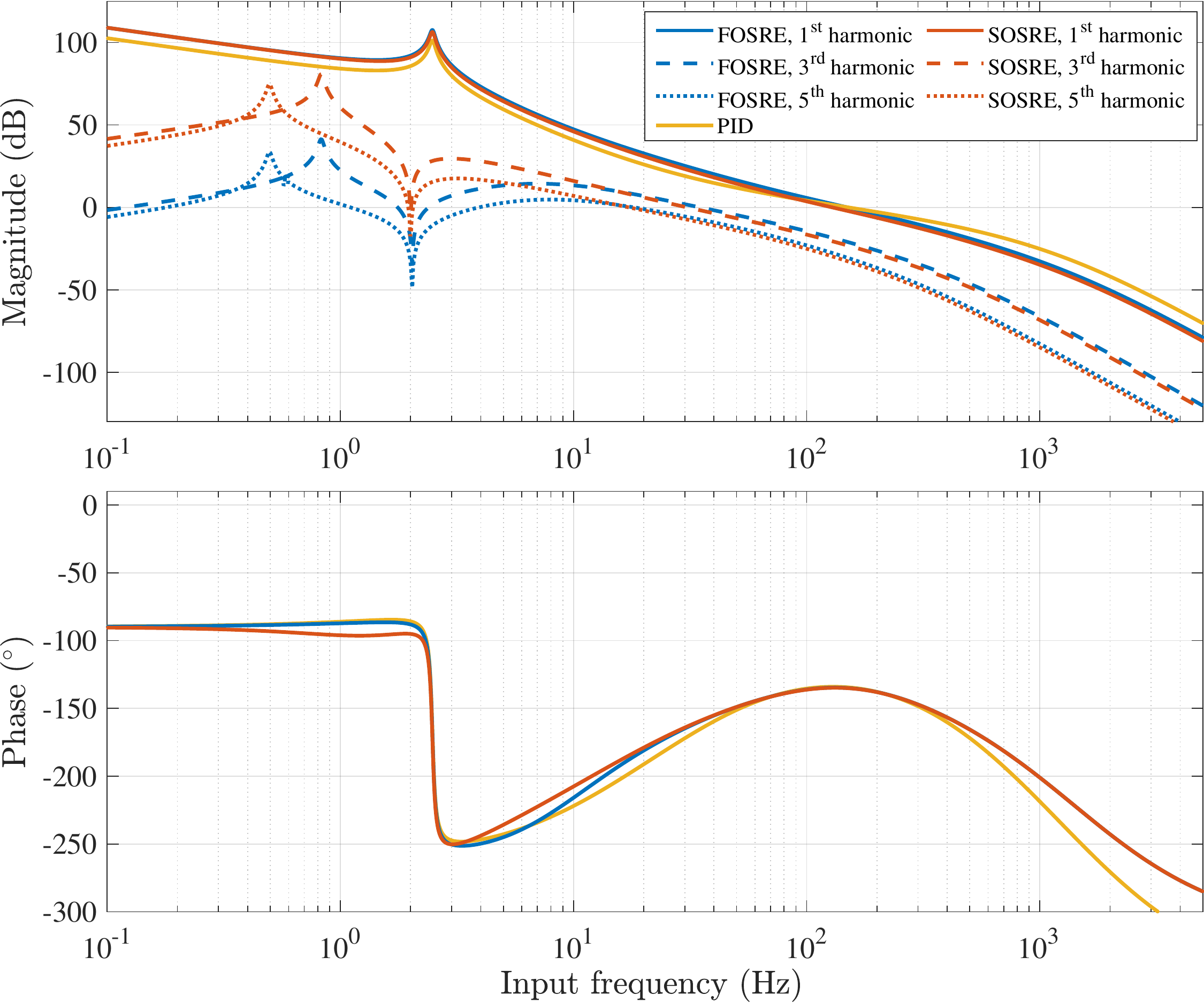}  
			\caption{Set No.1: Controllers designed to behave linear at 2 Hz}
			\label{fig:open_loop2}
		\end{subfigure}
		\begin{subfigure}{.5\textwidth}
			\centering
			% include second image
			\includegraphics[width=0.95\linewidth]{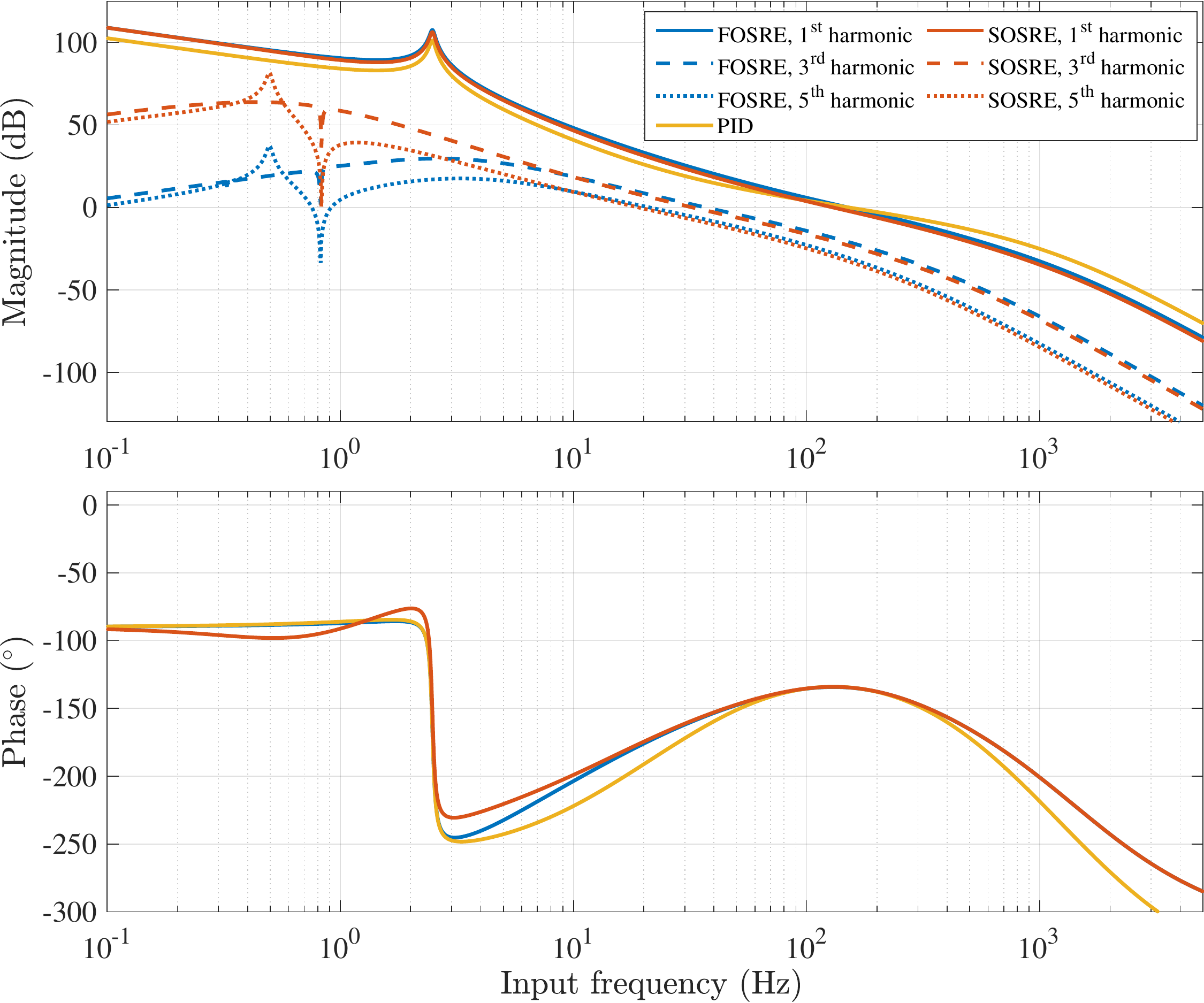}  
			\caption{Set No.2: Controllers designed to behave linear at 0.8 Hz to cancel $3^\text{rd}$ order harmonic peak}
			\label{fig:op2n_loop_0.8}
		\end{subfigure}
		\caption{HOSIDF of open loop for 3 systems designed based on a SOSRE CgLp, FOSRE CgLp and a PID. }
		\label{fig:open_loop}
	\end{figure*}
	\begin{figure*}[t!]
		\begin{subfigure}{.5\textwidth}
			\centering
			% include first image
			\includegraphics[width=0.95\linewidth]{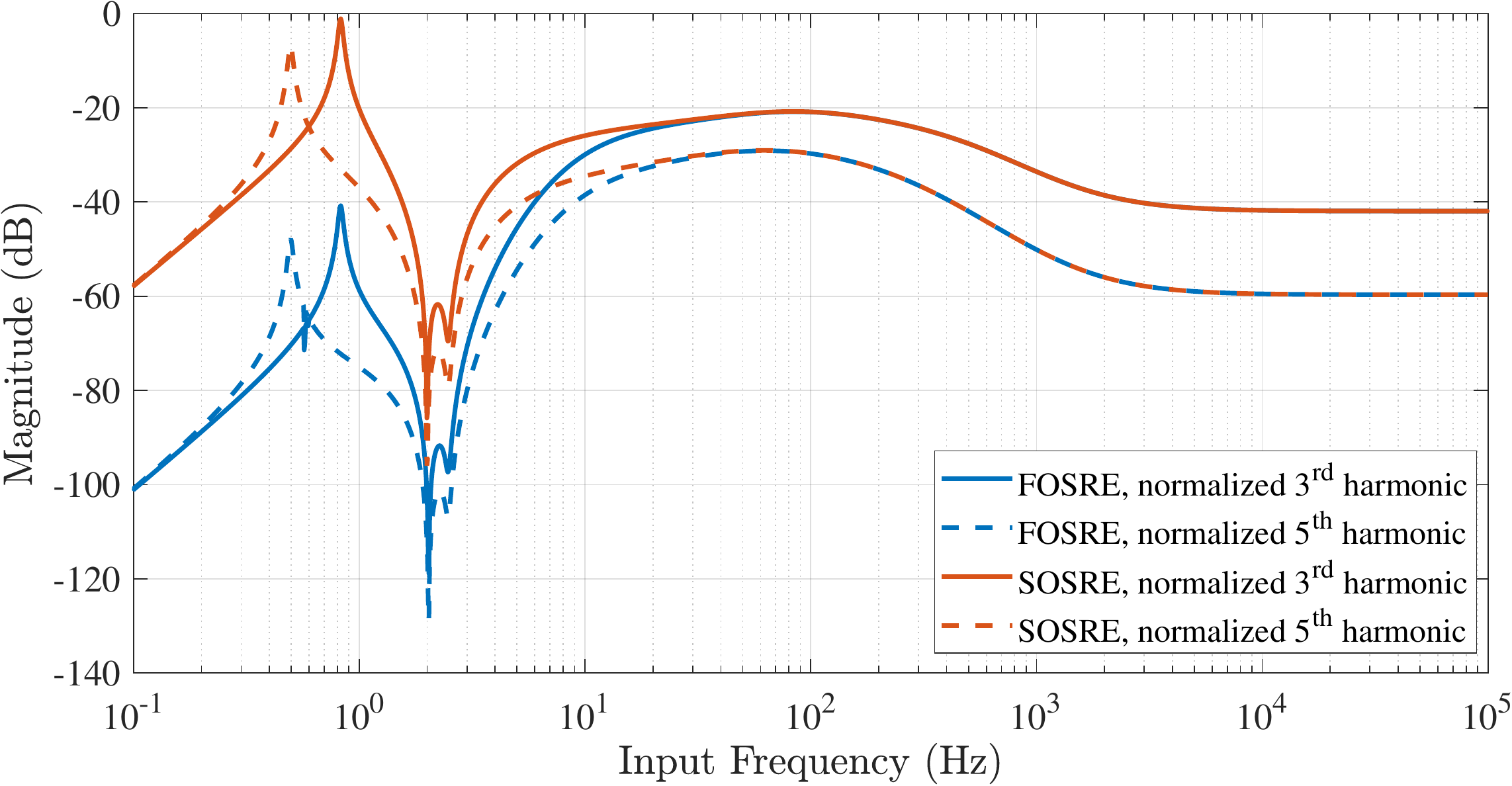}  
			\caption{Set No.1: Controllers designed to behave linear at 2 Hz}
			\label{fig:open_loop2_norm}
		\end{subfigure}
		\begin{subfigure}{.5\textwidth}
			\centering
			% include second image
			\includegraphics[width=0.95\linewidth]{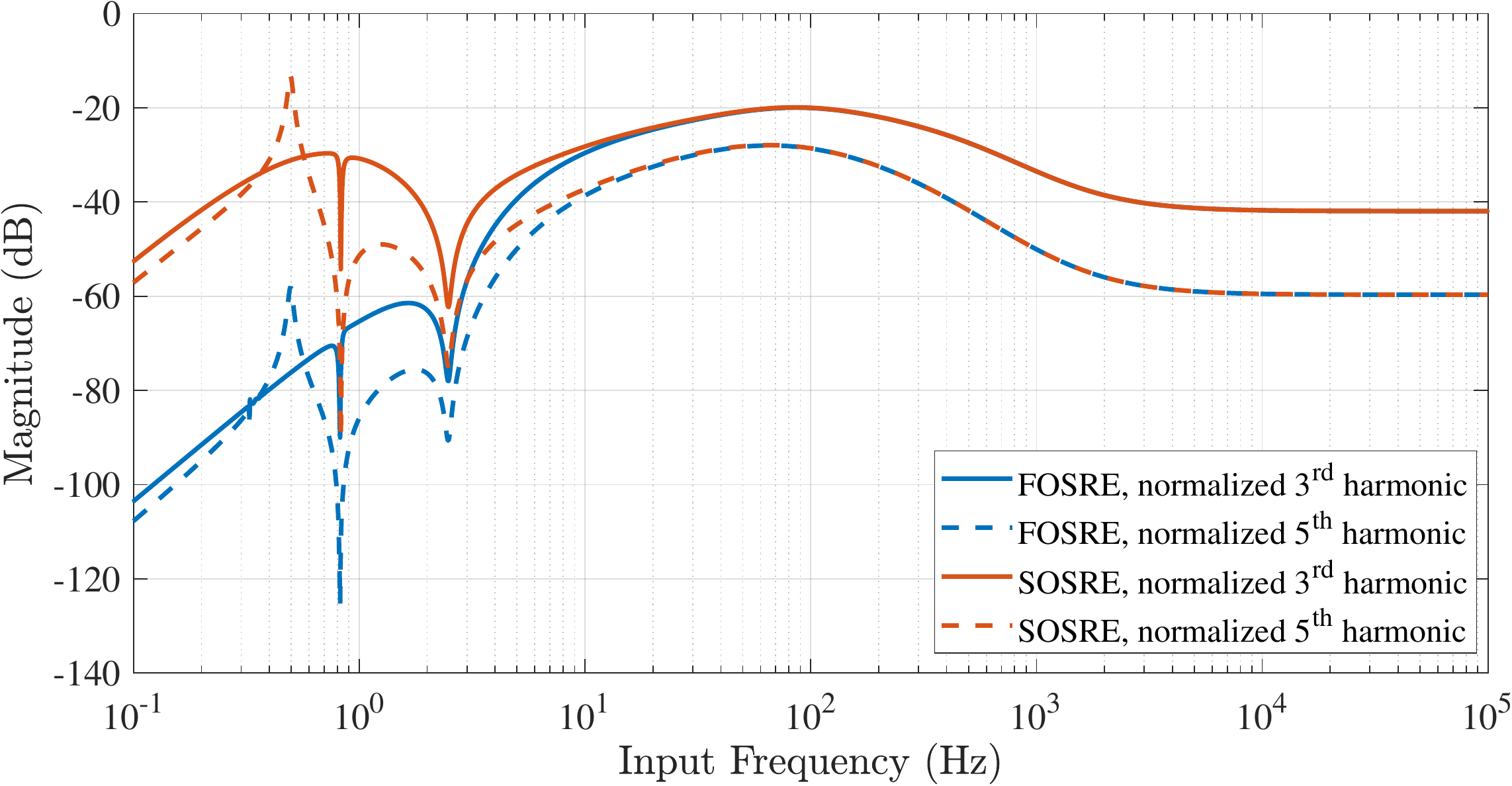}  
			\caption{Set No.2: Controllers designed to behave linear at 0.8 Hz to cancel $3^\text{rd}$ order harmonic peak}
			\label{fig:op2n_loop_0.8_norm}
		\end{subfigure}
		\caption{Normalized magnitude of higher-order harmonics with respect to first-order one for FOSRE and SOSRE.}
		\label{fig:open_loop_norm}
	\end{figure*}
	\subsection{Tuning guidelines}
	Tuning of FOSRE can be done through optimisation or several iteration of trial and error. The first parameter to choose is $\omega_{lb}$. As aforementioned this frequency can be a working frequency of the system or peak of the $3^\text{rd}$ harmonic. The cost function to minimize is $\psi(\omega)$ in the range of lower frequencies till $\omega_{lb}$. Denoting cross-over frequency as $\omega_c$, one can follow the following steps as a rule of thumb to achieve a favourable configuration:
	\begin{enumerate}
		\item \hl{\textbf{Choose} $\omega_{lb}$.}
		\item \hl{\textbf{Set} $\lambda$ \textbf{to be} -0.1\textbf{.}}
		\item \hl{\textbf{Set} $\beta$ \textbf{to be} 1\textbf{.}}
		\item \hl{\textbf{Optimise} $\omega_{l}$ \textbf{and} $\omega_{r \alpha}$ \textbf{to minimise} $\left|\psi(\omega)\right|$ \textbf{at frequencies lower than} $\omega_{lb}$\textbf{, constrained to} $\psi(\omega_{lb})=0$. }
		\item \hl{\textbf{Is} $\psi(\omega_{c})<-85^\circ$\textbf{? If yes, proceed to 7, if not, decrease} $\beta$ \textbf{by} 0.1.}
		\item \hl{\textbf{Is} $\beta=0$\textbf{? if yes, decrease} $\lambda$ \textbf{by} -0.1 \textbf{and return to 3, if not, return to 4.}}
		\item \hl{\textbf{Choose} $\gamma$ \textbf{in} $[-1\quad 1)$ \textbf{to achieve the phase margin required.}}
	\end{enumerate} 
	\section{An illustrative example}   
	In order to validate the increase in performance of the system in terms of steady-state tracking by suppressing the higher-order harmonics, three controllers were designed and studied in simulation. This section presents the results of the comparison of a FOSRE CgLp with a SOSRE CgLp and a PID.
	\subsection{Plant}
	The plant which is simulated is a custom-designed precision stage that is actuated with the use of a Lorentz actuator. This stage is linear-guided using two flexures to attach the Lorentz actuator to the base of the stage and actuated at the centre of the flexures. With a laser encoder the position of the fine stage is read out with 10nm resolution. A picture of the setup can be found in Fig.~\ref{fig:setup}. The identified transfer function for the plant is:
	\begin{equation}
	G(s)=\frac{3.038e4}{{{s}^{2}}+0.7413s+243.3}.
	\label{eq:plant}
	\end{equation}
	This plant has a relatively high resonance peak around 2.5 Hz which will cause high peaks in higher-order harmonics in frequencies below 1 Hz.
	\begin{figure*}[t!]
		\begin{subfigure}{.5\textwidth}
			\centering
			% include first image
			\includegraphics[width=0.9\linewidth]{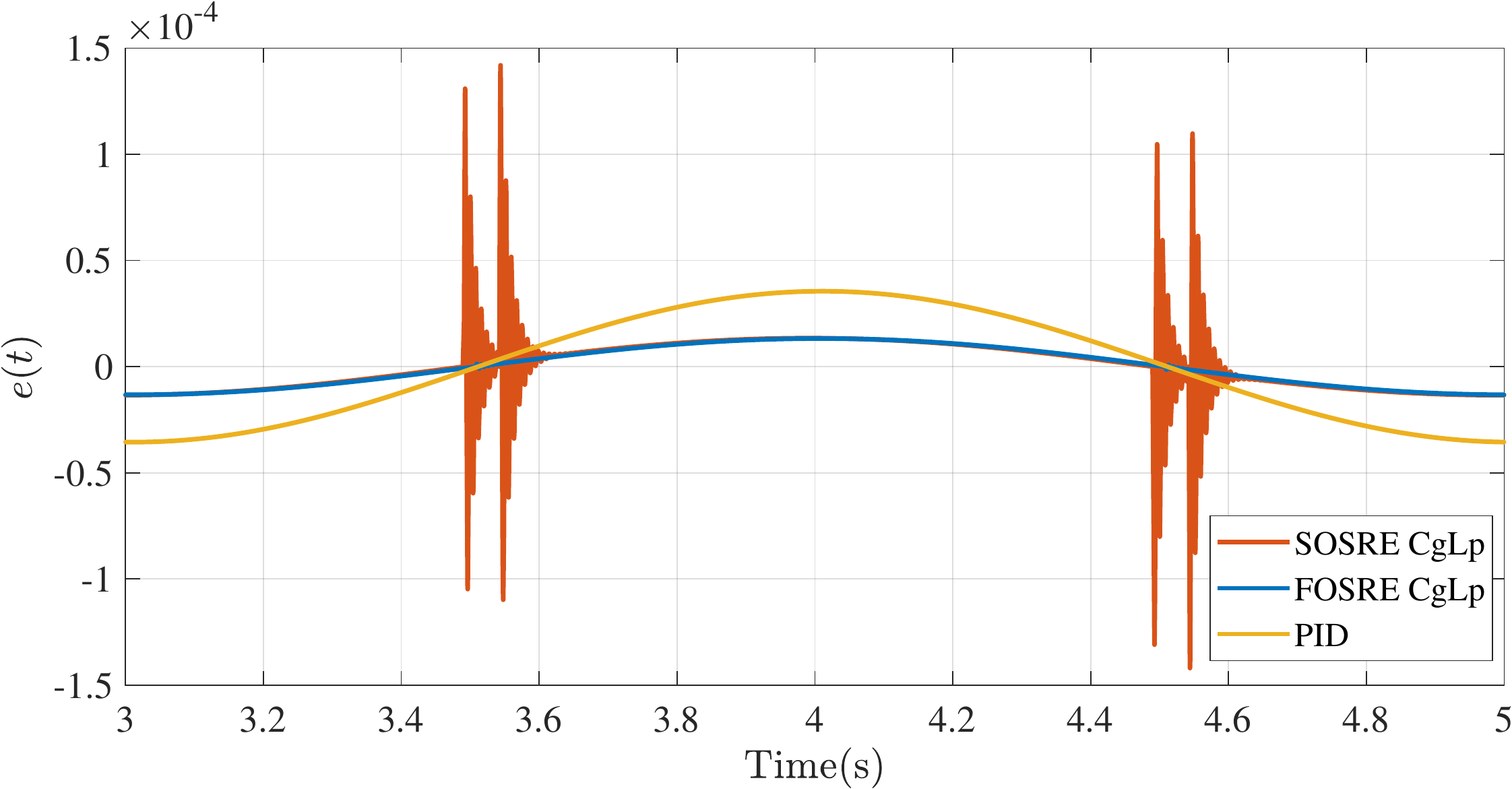}  
			\caption{Systems in set No. 1, $r(t)=\sin(\pi t)$}
			\label{fig:2_0.5}
		\end{subfigure}
		\begin{subfigure}{.5\textwidth}
			\centering
			% include second image
			\includegraphics[width=0.9\linewidth]{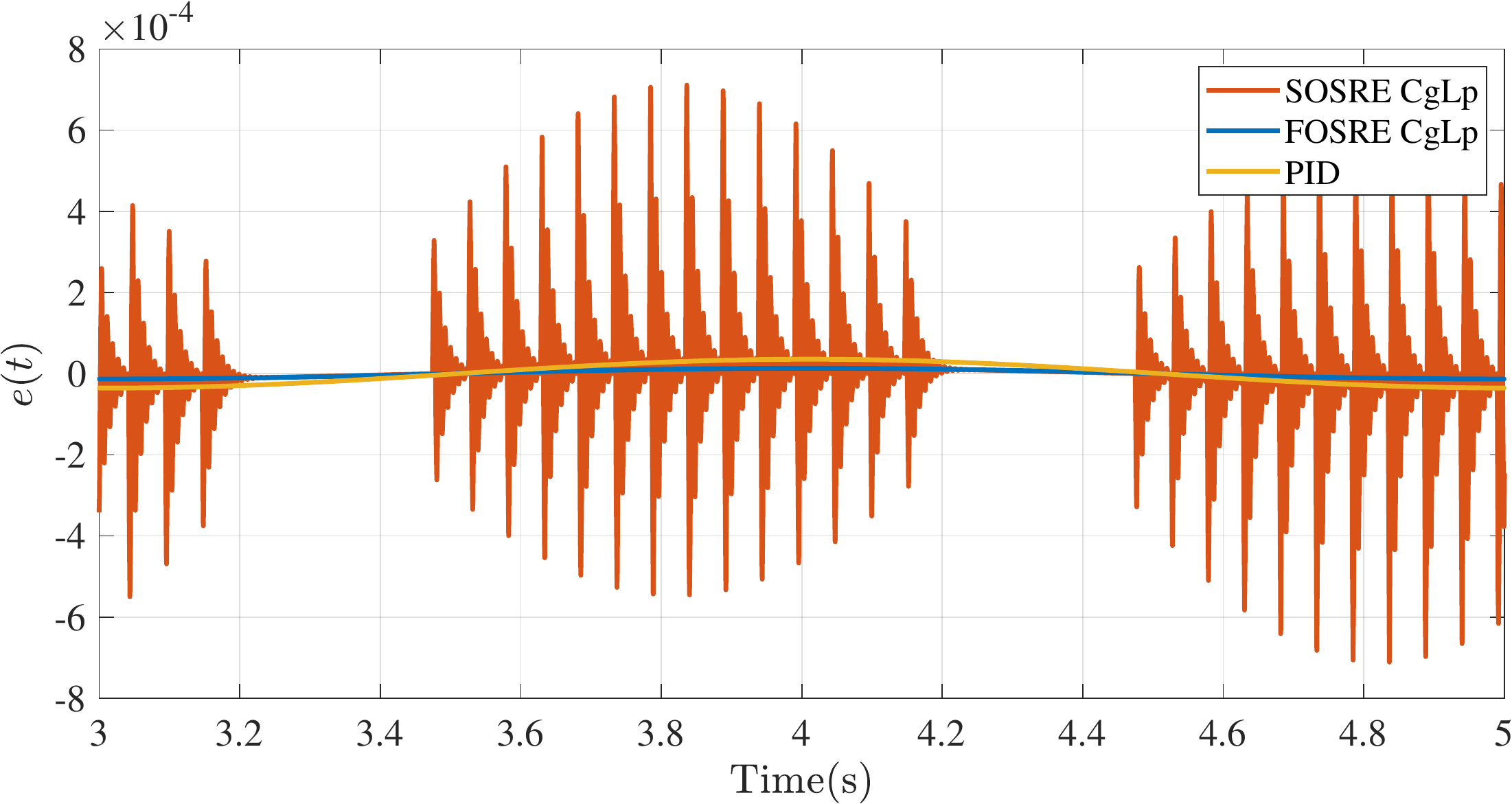}  
			\caption{Systems in set No. 2, $r(t)=\sin(\pi t)$}
			\label{fig:0.8_0.5}
		\end{subfigure}
		
		~\\
		
		\begin{subfigure}{.5\textwidth}
			\centering
			% include third image
			\includegraphics[width=0.9\linewidth]{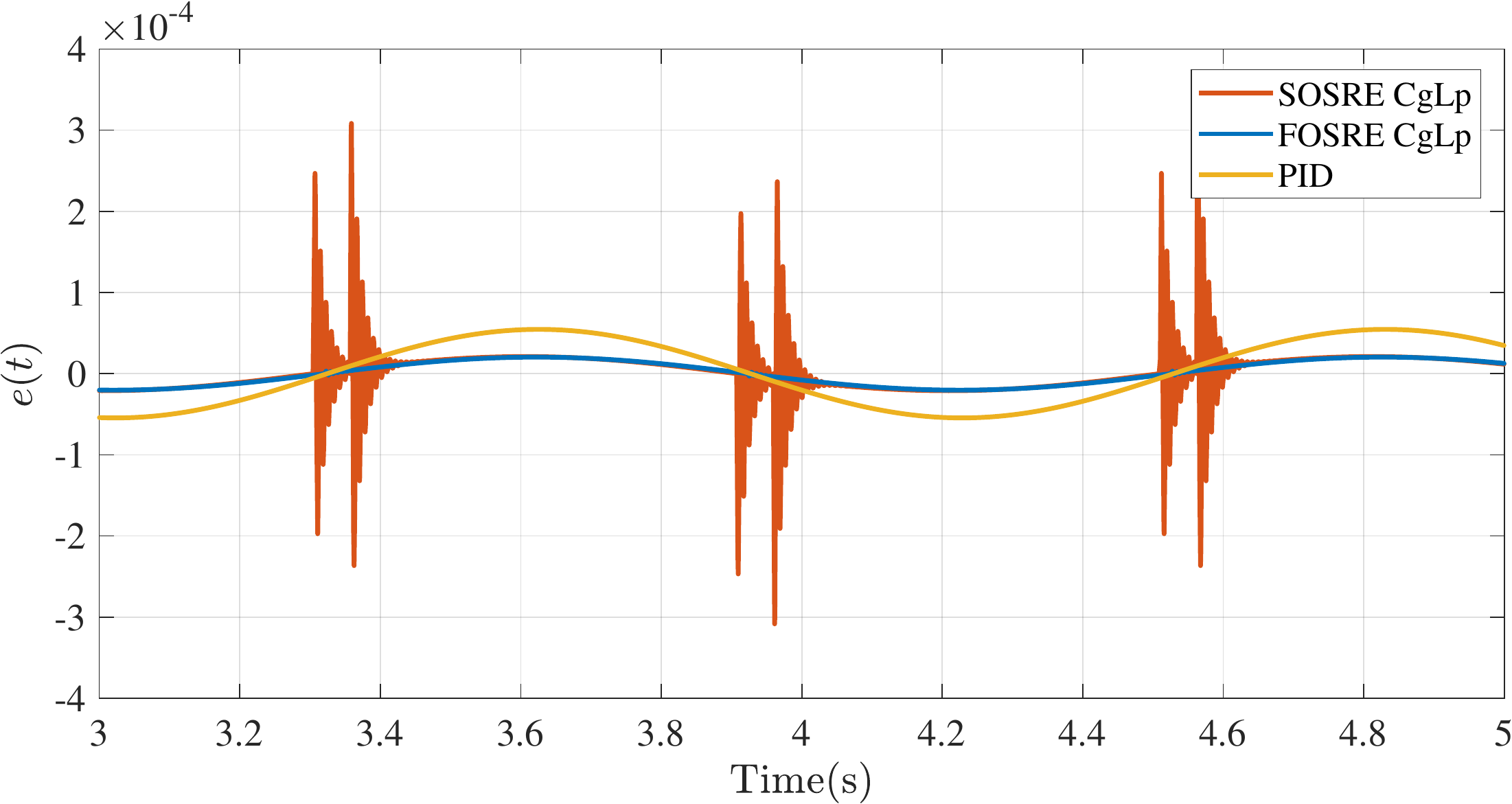}  
			\caption{Systems in set No. 1, $r(t)=\sin(1.6\pi t)$}
			\label{fig:2_0.8}
		\end{subfigure}
		\begin{subfigure}{.5\textwidth}
			\centering
			% include fourth image
			\includegraphics[width=0.9\linewidth]{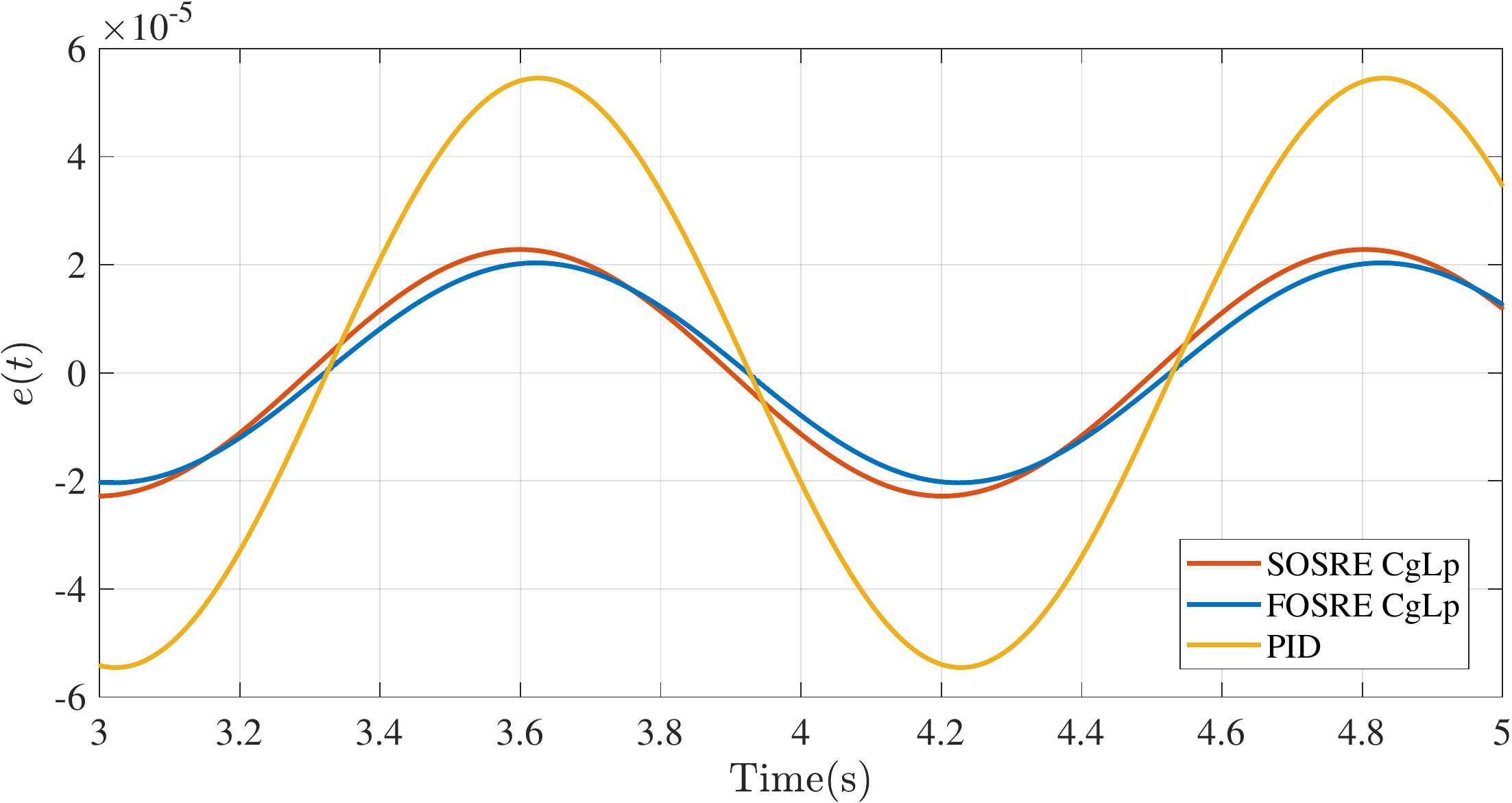}  
			\caption{Systems in set No. 2, $r(t)=\sin(1.6\pi t)$}
			\label{fig:0.8_0.8}
		\end{subfigure}
		\begin{subfigure}{.5\textwidth}
			\centering
			% include first image
			\includegraphics[width=0.9\linewidth]{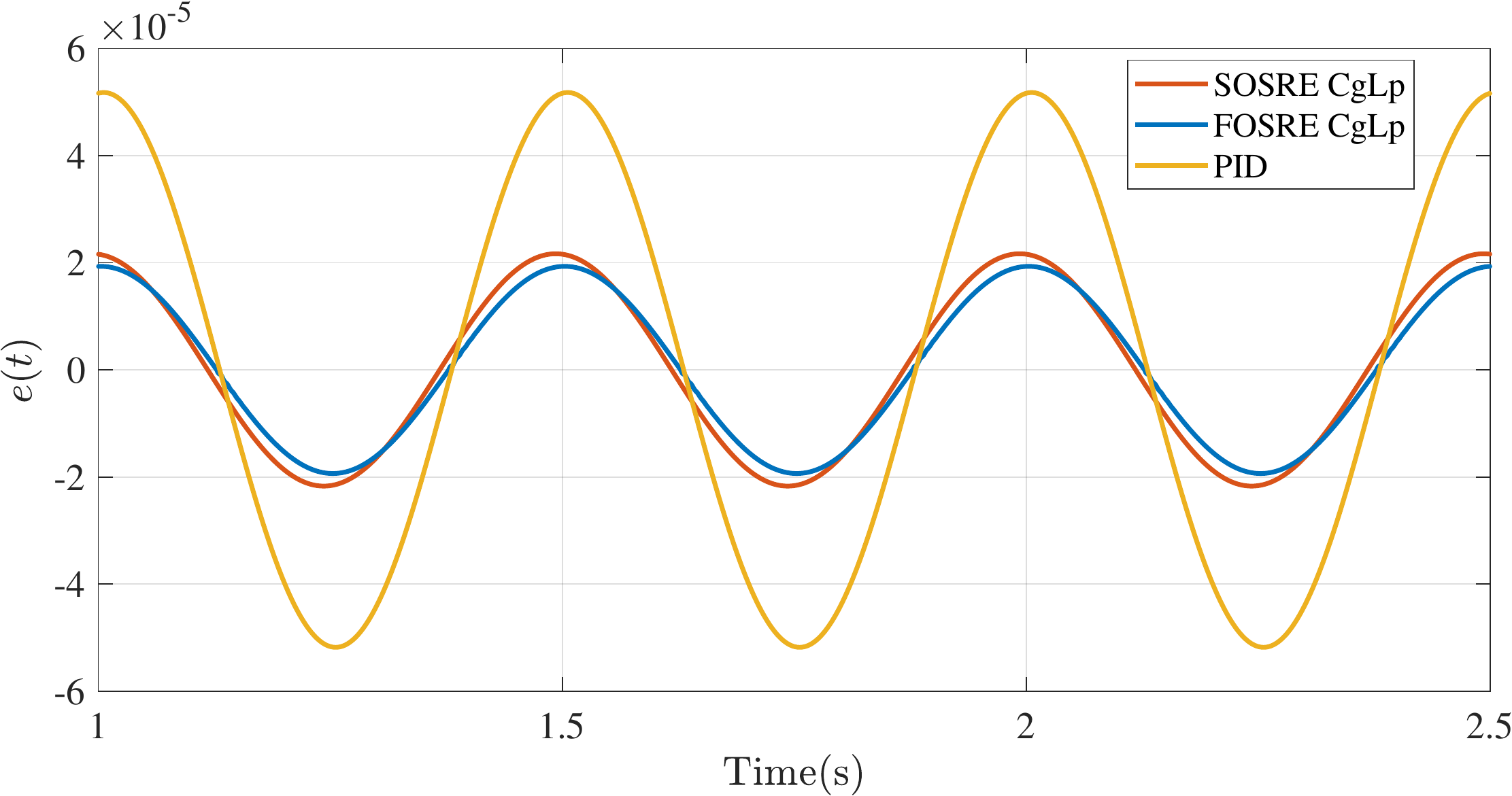}  
			\caption{Systems in set No. 1, $r(t)=\sin(4\pi t)$}
			\label{fig:2_2}
		\end{subfigure}
		\begin{subfigure}{.5\textwidth}
			\centering
			% include second image
			\includegraphics[width=0.9\linewidth]{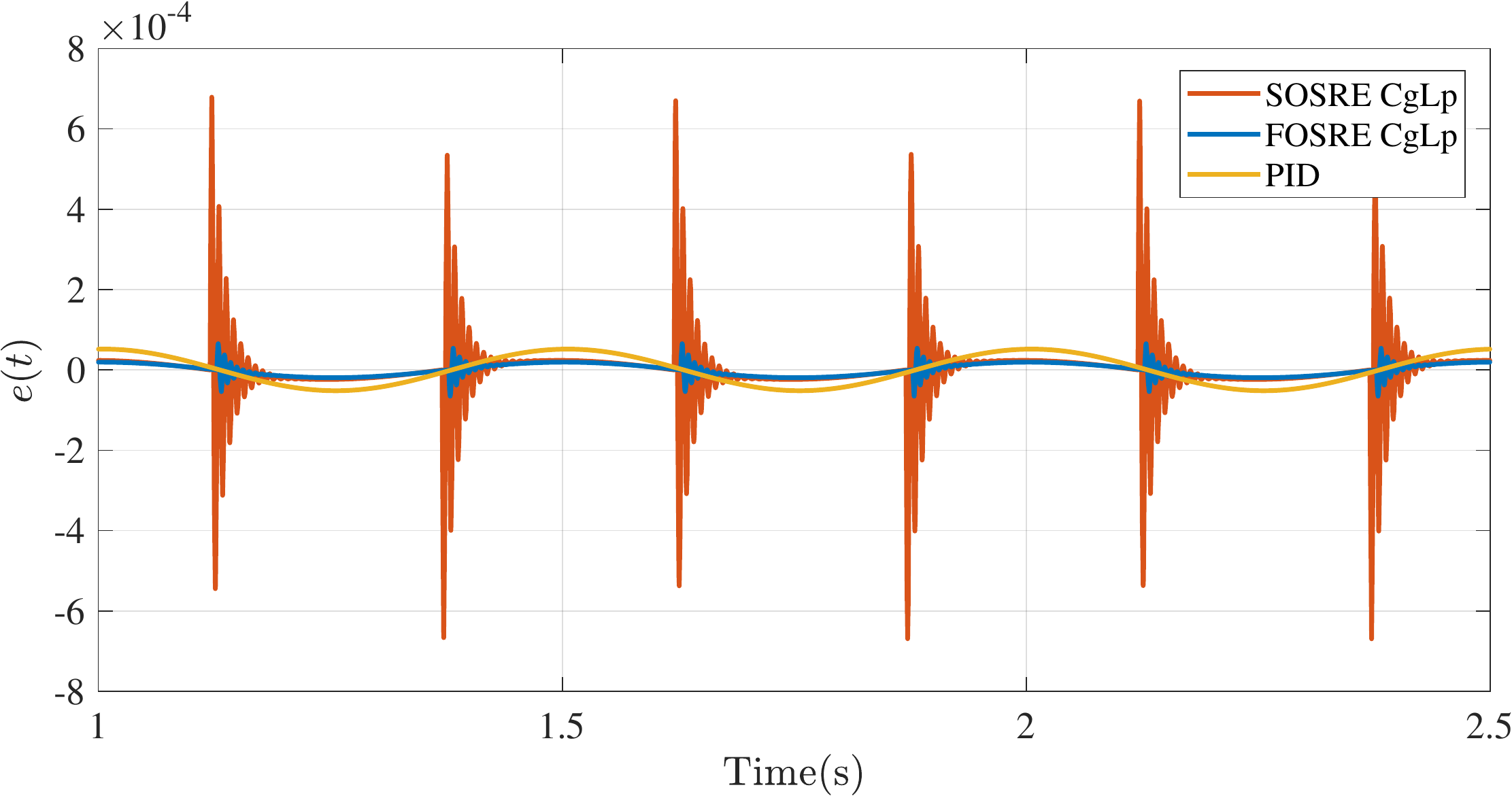}  
			\caption{Systems in set No. 2, $r(t)=\sin(4\pi t)$}
			\label{fig:0.8_2}
		\end{subfigure}
		
		~\\
		
		\begin{subfigure}{.5\textwidth}
			\centering
			% include third image
			\includegraphics[width=0.9\linewidth]{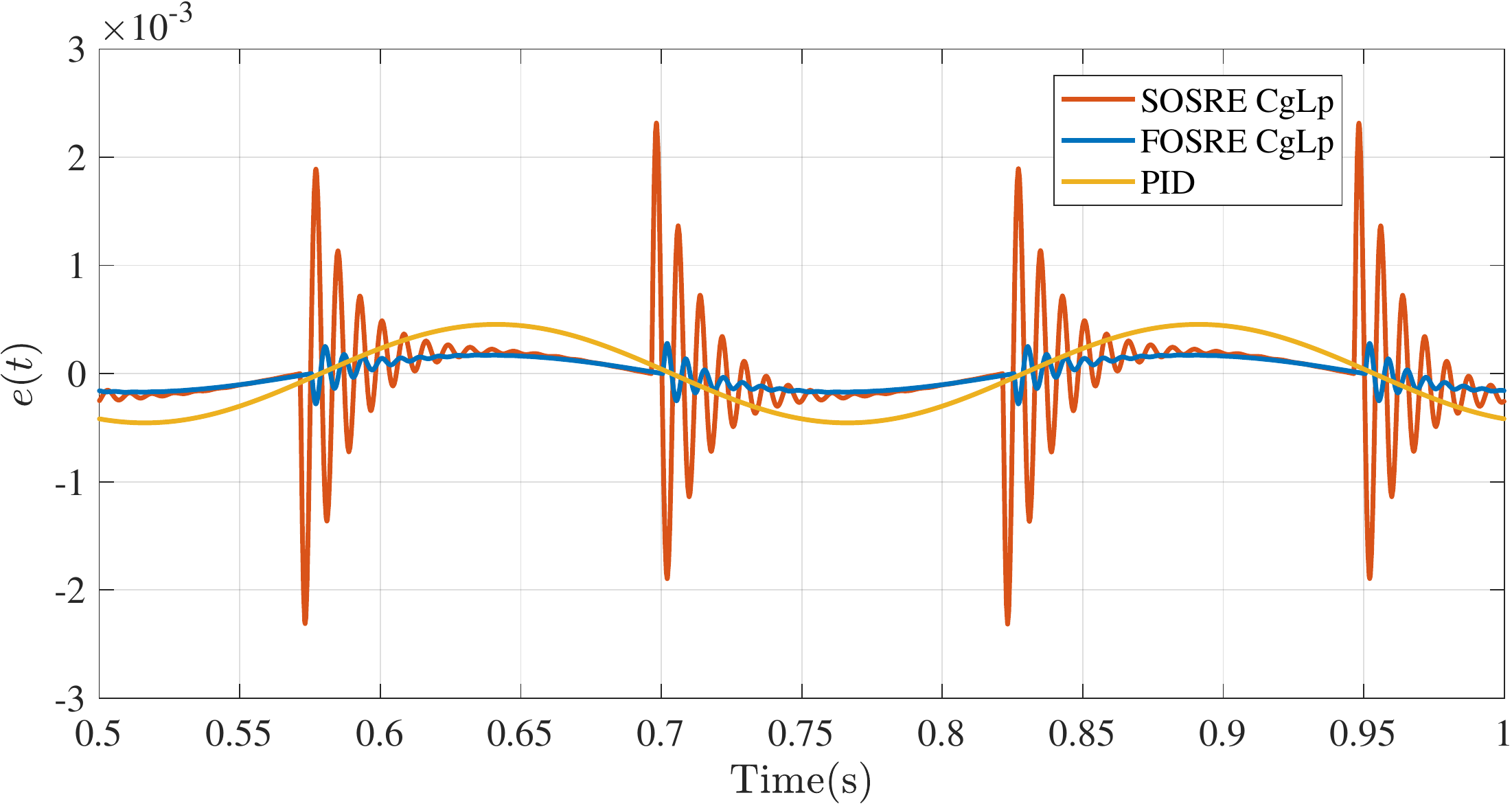}  
			\caption{Systems in set No. 1, $r(t)=\sin(8\pi t)$}
			\label{fig:2_4}
		\end{subfigure}
		\begin{subfigure}{.5\textwidth}
			\centering
			% include fourth image
			\includegraphics[width=0.9\linewidth]{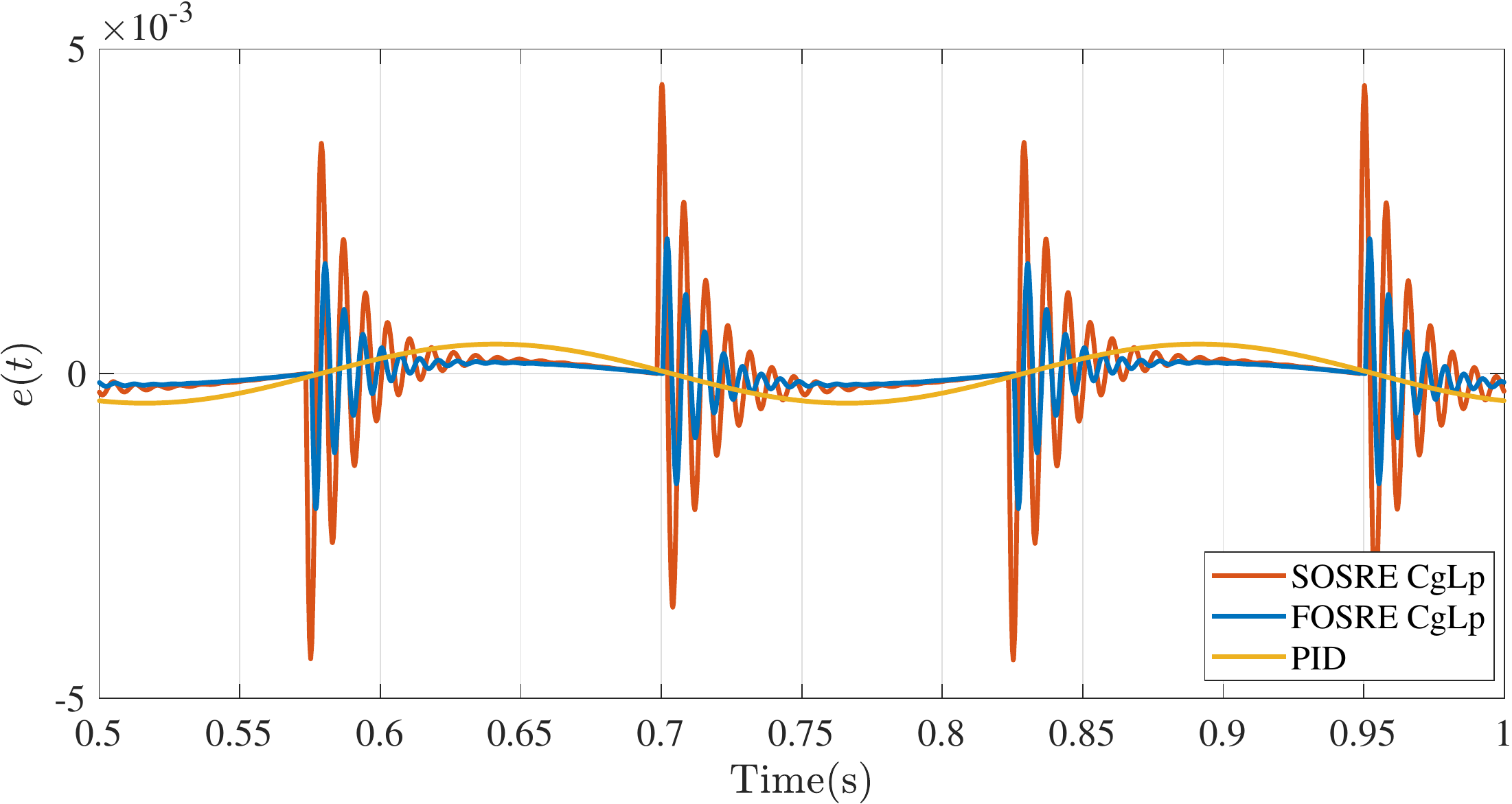}  
			\caption{Systems in set No. 2, $r(t)=\sin(8\pi t)$}
			\label{fig:0.8_4}
		\end{subfigure}
		\caption{The steady-state error of designed systems for tacking sinusoidal inputs.}
		\label{fig:simulation_error}
	\end{figure*}
	\subsection{Controller design approach}
\begin{figure}[t!]
	\centering
	\includegraphics[width=\columnwidth]{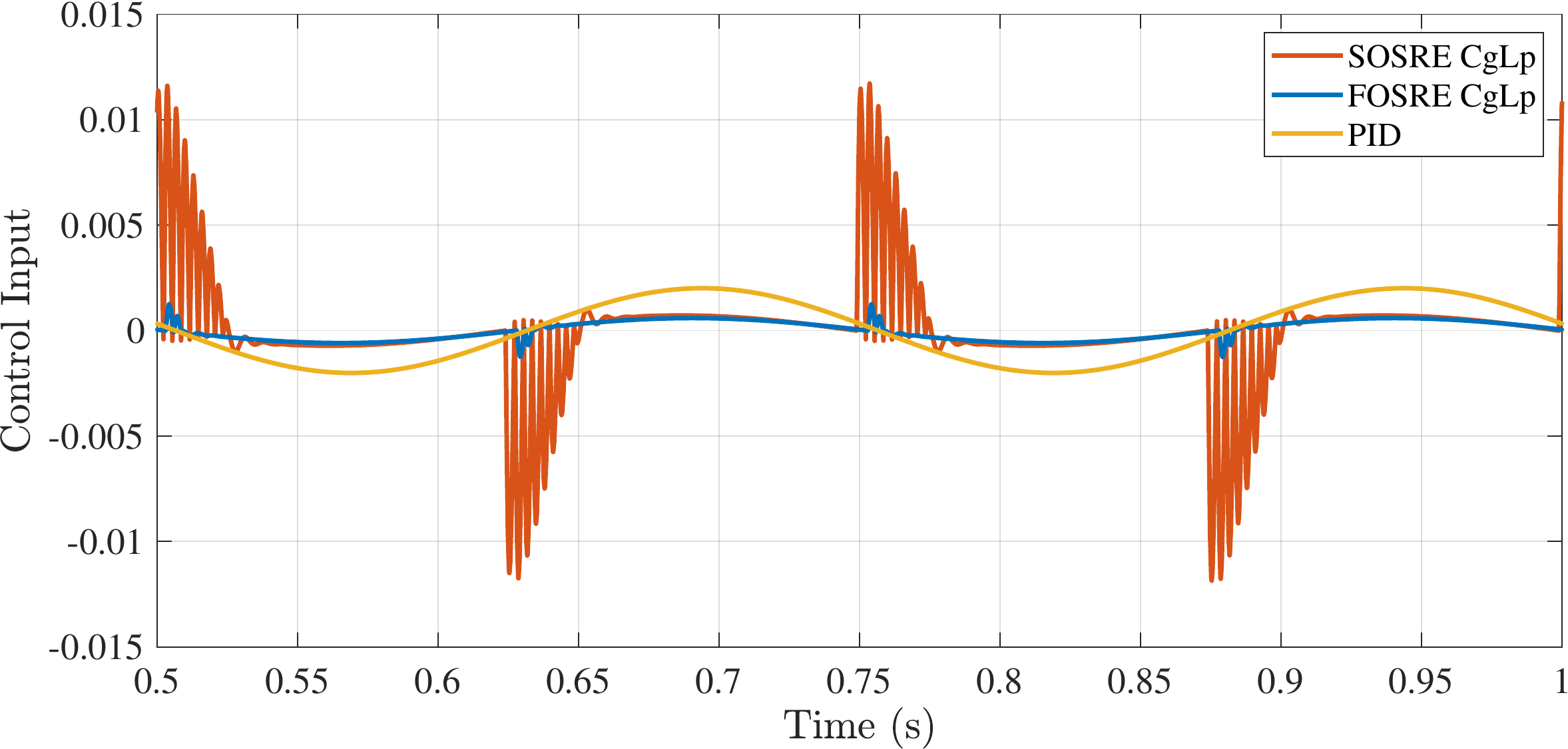}
	\caption{\hl{Control input of controllers in set No.1 for $r(t)=\sin(8\pi t)$.}}
	\label{fig:c_4_hz}
\end{figure}

\begin{figure}[t!]
	\centering
	\includegraphics[width=\columnwidth]{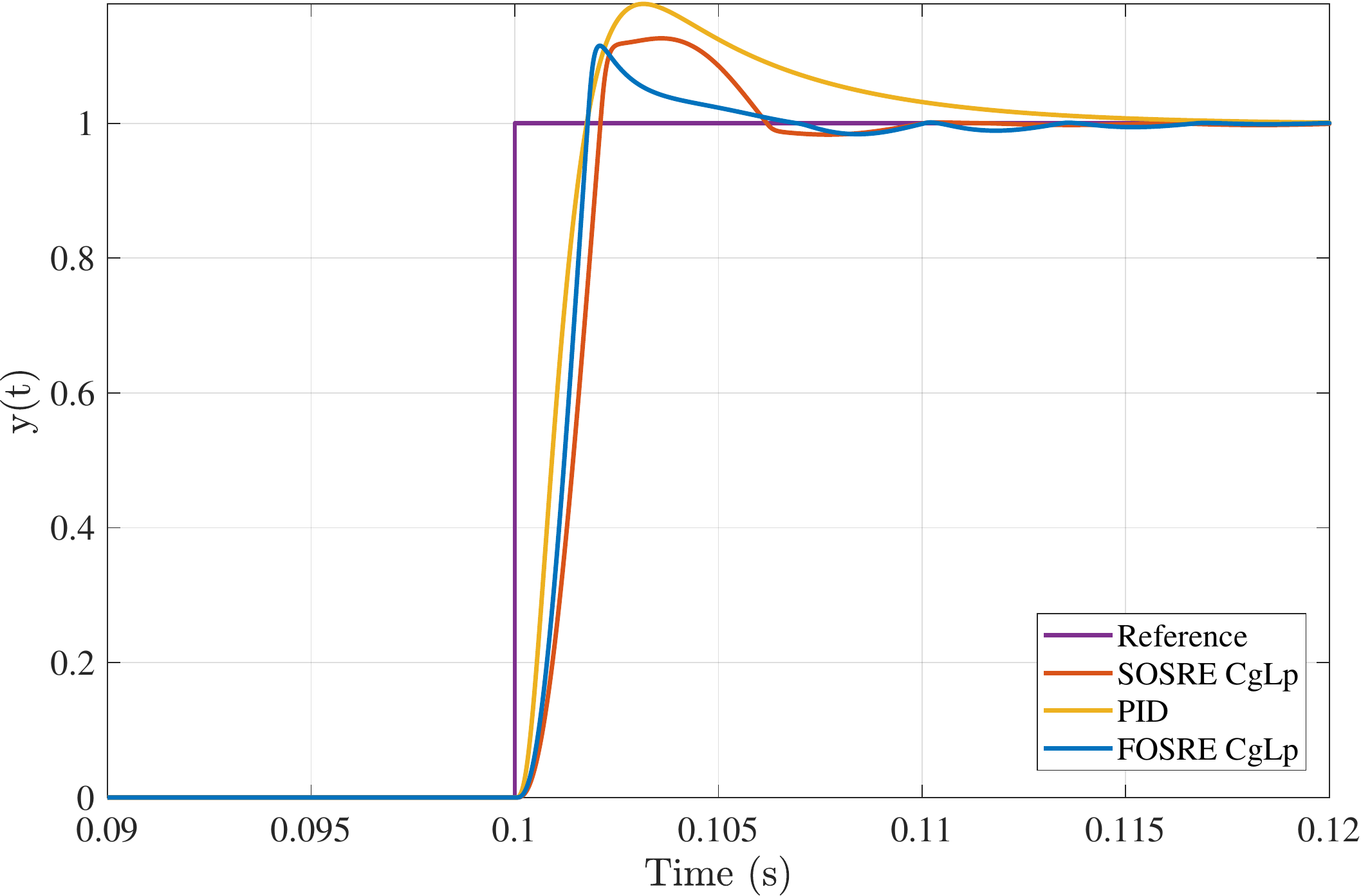}
	\caption{\hl{Step responce of controllers in set No. 1.}}
	\label{fig:step}
\end{figure}

	Two sets of controllers have been designed, each containing a FOSRE CgLp, a SOSRE CgLp and a PID. All of the controllers have designed for a bandwidth of 150 Hz and $45^\circ$ of phase margin considering a sensitivity peak below 6 dB criteria for robustness.\\
	%The PID is the same in both of the sets with $\omega_{i}=15$ Hz, $\omega_{d}=32$ Hz and $\omega_{t}=705$ Hz with a second-order low pass filter $\omega_{f}=1500$ Hz. 
	In set No. 1, assuming a main working frequency of 2~Hz, FOSRE CgLp and SOSRE CgLp are designed in a manner to have $\omega_{lb}=2$ Hz. This means the reset controllers will behave linearly in terms of steady-state output and will generate no higher-order harmonics at said frequency. In set No. 2, reset controllers are designed to have $\omega_{lb}=0.8$ Hz, which is the frequency of the peak of the third harmonic. Considering the discussion in Section~\ref{sec:parameters}, parameters of the FOSRE are chosen in a manner that $\psi$ stays as close as possible to zero in frequencies below $\omega_{lb}$. mentioned, such a design is not possible for SOSRE. Figure~\ref{fig:closed-loop} and Table~\ref{tab:params} show the closed-loop block diagram and the parameters for each controller.\\
	In order to be able to verify the stability of the reset systems through the so-called $H_\beta$ condition, a relatively weak derivate has been added to the design of CgLp's to provide a phase margin of $5^\circ$ for the base-linear system.  Thus, the CgLp's are providing the remaining $40^\circ$ required.\\
	The HOSIDF analysis of the open-loop of the designed system sets are presented in Fig.~\ref{fig:open_loop}. As expected, SOSRE CgLp's produce larger higher-order harmonics at low frequencies than the FOSRE ones, while they both have the same first-order DF and provide the same phase margin. The presence of large higher-order harmonics can jeopardise the tracking performance of the controller, as it invalidates the assumption of design based on DF.\\
	To better clarify the frequencies at which the tracking performance is weak and where it is the ideal, one can refer to normalised magnitude of higher-order harmonics with respect to first-order one in Fig.~\ref{fig:open_loop_norm}. According to this figure, one can predict the lower values of this plot indicates closer-to-ideal behaviour for CgLp. As a rule of thumb if the normalised magnitude if higher-order harmonics are below -40 dB, their effect is negligible provided that their magnitude is different enough to prevent their constructive behaviour from deteriorating the performance.\\
	Referring to Fig.~\ref{fig:open_loop} and~\ref{fig:open_loop_norm}, it is clearly shown that using FOSRE architecture in CgLp framework, higher-order harmonics can be suppressed at lower frequencies while maintaining them at the crossover frequency region to provide the required phase margin. Notice that the normalised higher-order harmonics are almost 0 dB at their peaks for SOSRE, which shows that they are almost equal to first-order one.\\
	Furthermore, it is shown that at $\omega_{lb}$, higher-order harmonics will be zero and this can be used to cancel out the higher-order harmonics peaks as shown in Fig~\ref{fig:op2n_loop_0.8}.\\
	To validate the performance of the controllers in closed-loop, a simulation has been performed in the Simulink environment of Matlab for tracking four sinusoidal waves of 0.5, 0.8, 2 and 4 Hz. The resulted error plots are depicted in Fig.~\ref{fig:simulation_error}. Furthermore, the Root Mean Square (RMS) and Integral Absolute Error (IAE) for steady-state error is presented in Table~\ref{tab:rms_iae_2} and ~\ref{tab:rms_iae_08}. The plots and tables verify that if higher-order harmonics are small enough, the reset controllers outperform the linear controller in terms of steady-state tracking error. A SOSRE CgLp designed for a system with high resonance peak can outperform  PID around at $\omega_{lb}$; however, it has difficulties at other frequencies. Nevertheless, FOSRE CgLp because of its much smaller higher-order harmonics has a much wider range of superiority, about 1.5 decades in this particular example.\\
	Since the direct relation of higher-order harmonics and tracking performance can be observed, it has to be noted that 0.5 and 0.8 Hz are the degenerate cases for reset controllers, however designing $\omega_{lb}$ to match and cancel out the peaks, the ideal tracking performance based on DF can be achieved. See Fig.~\ref{fig:0.8_0.8}.\\
	\hl{In order to compare the control actions of the designed controllers, the control input for controllers in set No.1 is depicted for a sinusoidal input of 4 Hz in Fig.}~\ref{fig:c_4_hz}.\hl{ Reset controllers are known for having large peaks in their control actions, however, according to Fig.}~\ref{fig:c_4_hz}\hl{, FOSRE CgLp shows smaller peaks due to reduced higher-order harmonics. The same holds for other frequencies and controllers in set No. 2, however they are not depicted for the sake of brevity.}\\
	\hl{At last, the step response for designed controllers in set No. 1 is shown in Fig.}~\ref{fig:step}. \hl{FOSRE and SOSRE CgLp show smaller overshoot and approximately the same settling time compared to PID. Controllers in set No. 2, show approximately the same step response.} 
% \usepackage{booktabs}
% \usepackage{graphicx}
% \usepackage{booktabs}
% \usepackage{graphicx}
% \usepackage{booktabs}
% \usepackage{multirow}
% \usepackage{graphicx}
% \usepackage{booktabs}
% \usepackage{graphicx}
% \usepackage{booktabs}
% \usepackage{graphicx}
\section{Conclusion}
This paper presented an architecture, named FOSRE, for reset elements based on a fractional-order integrator and the concept of having only one resetting integrator. It was shown that using this architecture in framework of CgLp; the higher-order harmonics can be suppressed at lower frequencies based on tuning the phase difference of input and output of the base linear system of the element. It was shown that at a particular frequency at which the mentioned phase difference is zero, no higher-order harmonics would be produced and the reset system would behave as a linear one in terms of steady-state output. Using this architecture, one can achieve the same phase margin as CgLp's introduced in the literature while increasing their tracking performance. The closed-loop performance of the FOSRE CgLp was compared with a SOSRE one and a PID in two different designs in simulation, and its superiority validated.\\
It was shown that one peak of higher-order harmonics could be cancelled out using the linear behaviour concept. However, this opens an opportunity for future researches on architectures which can behave linearly at more than one frequencies, as there are multiple peaks in higher-order harmonics.  As ongoing works, the performance of the FOSRE CgLp is being will be studied in the presence of noise and disturbance and afterwards in a practical setup. \\
% \usepackage{booktabs}
% \usepackage{graphicx}
\begin{table}[t!]
	\centering
	\caption{\hl{RMS and IAE of steady state error for controllers in set No. 1 for tracking the sinusoidal references of 0.5, 0.8, 2 and 4 Hz.}}
	\label{tab:rms_iae_2}
	\resizebox{\columnwidth}{!}{%
		\begin{tabular}{@{}l|cc|cc|cc|cc@{}}
			\toprule
			Frequency  & \multicolumn{2}{c|}{0.5 Hz} & \multicolumn{2}{c|}{0.8 Hz} & \multicolumn{2}{c|}{2 Hz} & \multicolumn{2}{c}{4 Hz} \\ \midrule
			Metric     & RMS          & IAE          & RMS          & IAE          & RMS         & IAE         & RMS         & IAE        \\ \midrule
			FOSRE CgLp & 9.36e-6      & 1.68e-5      & 1.47e-5      & 2.68e-5      & 1.36e-5     & 1.84e-5     & 1.31e-4     & 5.99e-5    \\
			SOSRE CgLp & 1.62e-5      & 2.22e-5      & 3.64e-5      & 4.30e-5      & 1.53e-5     & 2.07e-5     & 4.76e-4     & 1.44e-4    \\
			PID        & 2.51e-5      & 4.52e-5      & 3.96e-5      & 7.21e-5      & 3.66e-5     & 4.94e-5     & 3.21e-4     & 1.44e-4    \\ \bottomrule
		\end{tabular}%
	}
\end{table}
\begin{table}[t!]
	\centering
	\caption{\hl{RMS and IAE of steady state error for controllers in set No. 2 for tracking the sinusoidal references of 0.5, 0.8, 2 and 4 Hz.}}
	\label{tab:rms_iae_08}
	\resizebox{\columnwidth}{!}{%
		\begin{tabular}{@{}l|cc|cc|cc|cc@{}}
			\toprule
			Frequency  & \multicolumn{2}{c|}{0.5 Hz} & \multicolumn{2}{c|}{0.8 Hz} & \multicolumn{2}{c|}{2 Hz} & \multicolumn{2}{c}{4 Hz} \\ \midrule
			Metric     & RMS          & IAE          & RMS          & IAE          & RMS         & IAE         & RMS         & IAE        \\ \midrule
			FOSRE CgLp & 9.37e-6      & 1.68e-5      & 1.47e-5     & 2.68e-5     & 1.60e-5    & 2.11e-5    & 3.95e-4    & 1.15e-4   \\
			SOSRE CgLp & 1.47e-4      & 1.63e-4      & 1.64e-5     & 2.99e-5     & 9.34e-5    & 6.13e-5    & 8.72e-4    & 2.28e-4   \\
			PID        & 2.51e-5      & 4.52e-5      & 3.96e-5     & 7.21e-5     & 3.65e-5    & 4.94e-5    & 3.21e-4    & 1.44e-4   \\ \bottomrule
		\end{tabular}%
	}
\end{table}
\section*{Acknowledgements}    
This work was supported by NWO, through OTP TTW project \#16335.
\section*{Conflict of interest} The authors declare that they have no conflict of interest. 
	%\begin{table}
	%\begin{tabular}{lll}
	%\hline\noalign{\smallskip}
	%\end{tabular}
	%\end{table}

	%\begin{acknowledgements}
	%\end{acknowledgements}


	\bibliographystyle{unsrt}      % mathematics and physical sciences
	\bibliography{ref}   % name your BibTeX data base
	
	%\begin{thebibliography}{}
	%%
	%% and use \bibitem to create references. Consult the Instructions
	%% for authors for reference list style.
	%\end{thebibliography}
	
\end{document}